\documentclass[aps,twocolumn,showpacs,superscriptaddress]{revtex4-1}
\usepackage{amssymb}
\usepackage{amsmath}
\usepackage{graphicx}
\usepackage{hyperref} 
\usepackage{mathtools}

\DeclarePairedDelimiter\ket{\lvert}{\rangle}
\DeclarePairedDelimiterX\braket[2]{\langle}{\rangle}{#1\,\delimsize\vert\,\mathopen{}#2}

\usepackage{multirow}
\usepackage{array}
\newcolumntype{P}[1]{>{\centering\arraybackslash}p{#1}}
\newcolumntype{M}[1]{>{\centering\arraybackslash}m{#1}}
\hypersetup{colorlinks=true,
            linkcolor=blue,
            citecolor=blue,
            urlcolor=orange}
        
\usepackage[resetlabels,labeled]{multibib}
\newcites{App}{App Readings}
\usepackage[dvipsnames]{xcolor}

\begin{document}

\title{
Local magnetic moment formation and Kondo screening \\in the presence of Hund exchange: the two-band Hubbard model analysis}
\author{T. B. Mazitov}
\affiliation{Center for Photonics and 2D Materials, Moscow Institute of Physics and Technology, Institutsky lane 9, Dolgoprudny,
141700, Moscow region, Russia}
\author{A. A. Katanin}
\affiliation{Center for Photonics and 2D Materials, Moscow Institute of Physics and Technology, Institutsky lane 9, Dolgoprudny,
141700, Moscow region, Russia}
\affiliation{M. N. Mikheev Institute of Metal Physics, Kovalevskaya Street 18, 620219
Ekaterinburg, Russia.}

\begin{abstract}
We study formation and screening of local magnetic moments in the two-band Hubbard model in the presence of Hund exchange interaction using dynamic mean field theory approach. The characteristic temperatures of the formation, beginning and full screening of local magnetic moments are obtained from  the analysis of temperature dependence  of the orbital, charge, and spin susceptibilities, as well as the respective instantaneous correlation functions. At half filling we find the phase diagram, which is similar to the single-orbital case with wide region of formation of local magnetic moments below the orbital Kondo temperature. Similarly to the single-orbital case, with decreasing temperature the screening of local magnetic moments is preceded by appearance of fermionic quasiparticles. In the two-band case the quasiparticles are however present 
also in some temperature region above the Mott insulating phase. At finite doping we find broad regime of presence of local magnetic moments, which coexist with incoherently or coherently moving holes. We also find, similarly to the half filled case, finite temperature interval in the doped regime, when the fermionic quasiparticles are formed, but do not yet screen local magnetic moments.
\end{abstract}

\maketitle

\section{Introduction}

The concept of Hund metals (HM) was introduced \cite{Pnictides,Pnictides1,Hund1,Hund2,Hund3,Hund4,Hund5,Freezing,Delft, Medici} to distinguish the substances where strong electron correlations are induced by Hund exchange from those where Coulomb interaction yields proximity to Mott insulating (MI) phase. The Hund metal behavior is obtained in iron pnictides \cite{Pnictides,Hund2,Pnic1,Pnic2}, ruthenates \cite{Hund1x,Hund2x},  elemental iron \cite{OurIron,Hausoel,OurFe1,OurFeGamma}, as well as some other transition metals \cite{Hausoel,OurV}, and their compounds \cite{OurFeNi}. Some aspects of the physics of high-temperature superconductors can be also described by an effective two-band model, in which Hund exchange plays an important role \cite{High-Tc,High-Tc1}. 

It was suggested in Refs. \cite{Hund3,Delft} that 
Hund metals are characterised by the so called spin-orbital separation, which corresponds to freezing of orbital degrees of freedom below the orbital Kondo temperature $T_{K,{\rm orb}}$. The latter temperature is much higher, than the Kondo temperature for the spin degrees of freedom $T_K$. This separation gives a possibility for the so called spin freezing \cite{Freezing} to occur at sufficiently low temperatures when the orbital degrees of freedom are not active. The spin freezing yields formation of local magnetic moments, which are important for magnetic and thermodynamic properties of Hund metals. 

The number of previous studies have contrasted the Hund metals to multi-band systems in the proximity to Mott insulating state at integer fillings, different from half filling (the so called Hundness vs. Mottness behavior), see, e.g., Refs. \cite{Hund1,Hund3,Delft,Deng,Fanfarillo,Medici,Pnictides1,Hund2}. At the same time, one can conjecture that the HM state away from half filling 
inherits to some extent its properties from the proximity the Mott state at {\it half filling}. This conjecture is confirmed by studies of the doping dependence of electronic, charge, and magnetic properties in broad range of dopings (see, e.g., Refs. \cite{Medici,Hund1,Pnictides1,Hund3,Fanfarillo}). In particular, the Hund metal behavior occurs with doping in the same part of the phase diagram, which is Mott insulating at half filling, and enhanced on approaching half filling.

Since the properties of Mott transition at {\it half filling} in multi-orbital models are similar to those in the single-orbital model, the HM state finds some similarities to the proximity of Mott state in the {\it single-band system}. Indeed, the temperature scales of start of the formation of local magnetic moments and beginning of their screening
in single-band systems were actively discussed recently \cite{Our1,Our2,Toschi,Katsnelson,Toschi_new}, and it was found that the temperature of the beginning of formation of local magnetic moments appears to be much higher, than the temperature of the onset of the screening. The former temperature is therefore analogous to that for the orbital freezing in multi-band system.

Finding similarities and differences in the formation of local magnetic moments in single- and multi-band systems allows therefore for better understanding of the physical properties of Hund metals, especially in the most important regime of the {\it unscreened} local magnetic moments, which are expected to form in the temperature range $T_K<T<T_{K,{\rm orb}}$. In particular, this study may reveal peculiarities of the local magnetic moment formation in multi-band systems at half filling, as well as uncover the reasons of their stability with doping.

To emphasize various aspects of the similarity and differences of the Mott transition in the single-orbital model and Hund metal behavior in multi-orbital model, we consider in the present paper the phase diagrams of the two-orbital model in the presence of Hund exchange interaction. The important  effect of Hund coupling for reduction of the critical interaction and formation local magnetic moments was emphasized in this model at half filling in previous DMFT studies \cite{Pruschke,2band1,Anisimov}. Ref. \cite{Werner} has performed investigations of various instabilities at and away from half filling. The differences of Mottness and Hundness behavior were studied for this model away from half filling in Refs. \cite{2orb1,Fanfarillo}. 

In the present paper we concentrate on the evolution of local magnetic moments with temperature, interaction, and doping, starting from the proximity to the Mott transition at half filling. We show that the effect of Hund exchange is crucial for the properties of LMM in the considered two-orbital model. By studying variety of physical observables within the dynamical mean-field theory (DMFT), we find the temperatures of the start of the formation of local magnetic moments, appearing well-defined fermionic quasiparticles (QP), and the onset of screening of local magnetic moments. We show that these scales are essentially different and comprise the phase diagram at half filling, which is similar to that obtained previously for the single-orbital case. 

Yet, away from half filling we find that in contrast to the single-orbital model, the local magnetic moments exist also in the unscreened phase in a broad range of doping levels. This allows us to connect our observations in the vicinity of half filling to previous studies of the doped multi-band Hubbard model, see, e.g., Ref. \cite{Fanfarillo}. 


The plan of the paper is the following. In Sect. II we formulate the model and the physical observables we study; in Sect. III we consider the properties of the half filled system, and in Sect IV we consider the off half filled case. Finally, in Sect. V we present Conclusions.

\section{The model and considered observables}

We consider the two-band Hubbard model with Ising symmetry of Hund exchange on the square lattice ({the obtained results are however expected to be qualitatively applicable for an arbitrary density of states})

\begin{align} \label{eq:hubbard}
H &= \sum_{ij, m, \sigma}{(t_{ij} - \mu\delta_{ij}) c^+_{im \sigma} c_{jm \sigma}}+U\sum_{im}{n_{im\uparrow}n_{im\downarrow}} \\&+ \sum_{i,m < {m'}, \sigma {\sigma'}}{({U'} -J\delta_{\sigma {\sigma'}})n_{im \sigma}n_{i{m'}{\sigma'}}}, \notag
\end{align}
where $c^\dagger_{i m \sigma} (c_{i m \sigma})$ is the electron creation (destruction) operator at site $i$, orbital $m = 1, 2$ and spin $\sigma, {\sigma'} = \uparrow,\downarrow$; $n_{i m \sigma} = c^\dagger_{im \sigma} c_{i m \sigma}$ is the electron number operator, $\mu$ is the chemical potential, and $t_{ij}=-t$ for the nearest neighbor sites. We assume the interorbital Coulomb interaction $U'=U-2J$. As the unit of energy we use the half bandwidth $D=4t=1$. In the majour part of calculations (if not stated differently) we fix the value of Hund exchange interaction $J=U/4$, which roughly corresponds to the Hund exchange in transition metals. 

We use the dynamic mean field theory \cite{DMFT}, by mapping the lattice model to the quantum impurity model, subject to the self-consistency condition. {It was shown previously for single-band Hubbard model \cite{Our1,Our2,Comment,LocDMFT1,ToschiAnd1} that near Mott metal-insulator transition DMFT describes formation of local magnetic moments and their screening. Using DMFT for the two-orbital model allows us to describe both, spin and orbital Kondo physics (see also Refs. \cite{Delft,Hund3,Deng,Comment,Medici,Fanfarillo}), which formally occurs due to hybridization of the impurity with the reservoir in the presence of the on-site Coulomb repulsion.}

To trace the formation of local moments we calculate in the self-consistent solution of DMFT the local spin susceptibility,
\begin{equation}
\chi_s(T) = \int_0^\beta \left\langle S_{i}^z(\tau) S_{i}^z(0) \right\rangle d\tau,
\end{equation}
where $S^z_i=S^z_{i,1}+S^z_{i,2}$, $S_{i,m}^z=(1/2)(c^+_{im\uparrow}c_{im\uparrow}-c^+_{im\downarrow}c_{im\downarrow})$ is the spin projection at the site i and imaginary time $\tau$, $\beta=1/T$ (Boltzmann's constant is put to unity). We also consider 
the local static charge susceptibility (local charge compressibility) 
\begin{equation} 
\chi_c(T)=\frac{dn}{d\mu}=\int_0^\beta \left( \left\langle n_i(\tau) n_i(0) \right\rangle - n^2 \right) d \tau,\label{ChargeCorr}
\end{equation} 
where $n_i(\tau) = \sum_m n_{i, m}(\tau)$, $n_{i, m}(\tau)=\sum_{\sigma} n_{im\sigma}$, and the change of the chemical potential $d\mu$ acts only at the impurity site, the static orbital susceptibility $\chi_{\rm orb}(T)\equiv\chi_{\rm orb}(T,0)$, where
\begin{align} \label{eq:chi_orb}
\chi_{\rm orb}(T,i\omega_n)=\frac{1}{4}\int_0^\beta &\langle (n_{i,1}(\tau) - n_{i,2}(\tau)) \notag\\&\times(n_{i,1}(0) - n_{i,2}(0)) \rangle e^{i\omega_n \tau} d \tau,
\end{align} 
is the dynamic orbital susceptibility ($\omega_n$ are the bosonic Matsubara frequencies), as well as the square of the instantaneous local magnetic moment
\begin{align} \label{eq:s2}
\left\langle {\mathbf S}_i^2 \right\rangle&=\left\langle ({\mathbf S}_{i,1}+{\mathbf S}_{i,2})^2 \right\rangle \notag\\&= 3n/4 - 3 \left\langle n_{im\uparrow} n_{im \downarrow} \right\rangle + 2 \left\langle S_{i,1}^z S_{i,2}^z \right\rangle 
\end{align}
{where we neglect correlations of transverse components of spins at different orbitals.}
Apart from that, we study the frozen spin ratio \cite{Freezing}
\begin{equation} 
R_s = \frac{\left\langle S_i^z(\beta/2) S_i^z(0) \right\rangle }{\left\langle (S_i^z)^2  \right\rangle},
\label{Froz}
\end{equation} 
which characterizes the degree of the local magnetic moment formation.

For computations, we use the hybridization expansion {continuous-time} quantum Monte Carlo \cite{cthyb1, cthyb2, cthyb3, cthyb4, cthyb5} impurity solver, implemented in the 
iQIST software package \cite{iQIST,iQISTNote}.
The analytical continuation of the fermionic and bosonic spectral functions is performed using {\it anacont} package \cite{ana_cont}.

\section{Results at half filling}

\begin{figure}[t]
\includegraphics[width=1.0\linewidth]{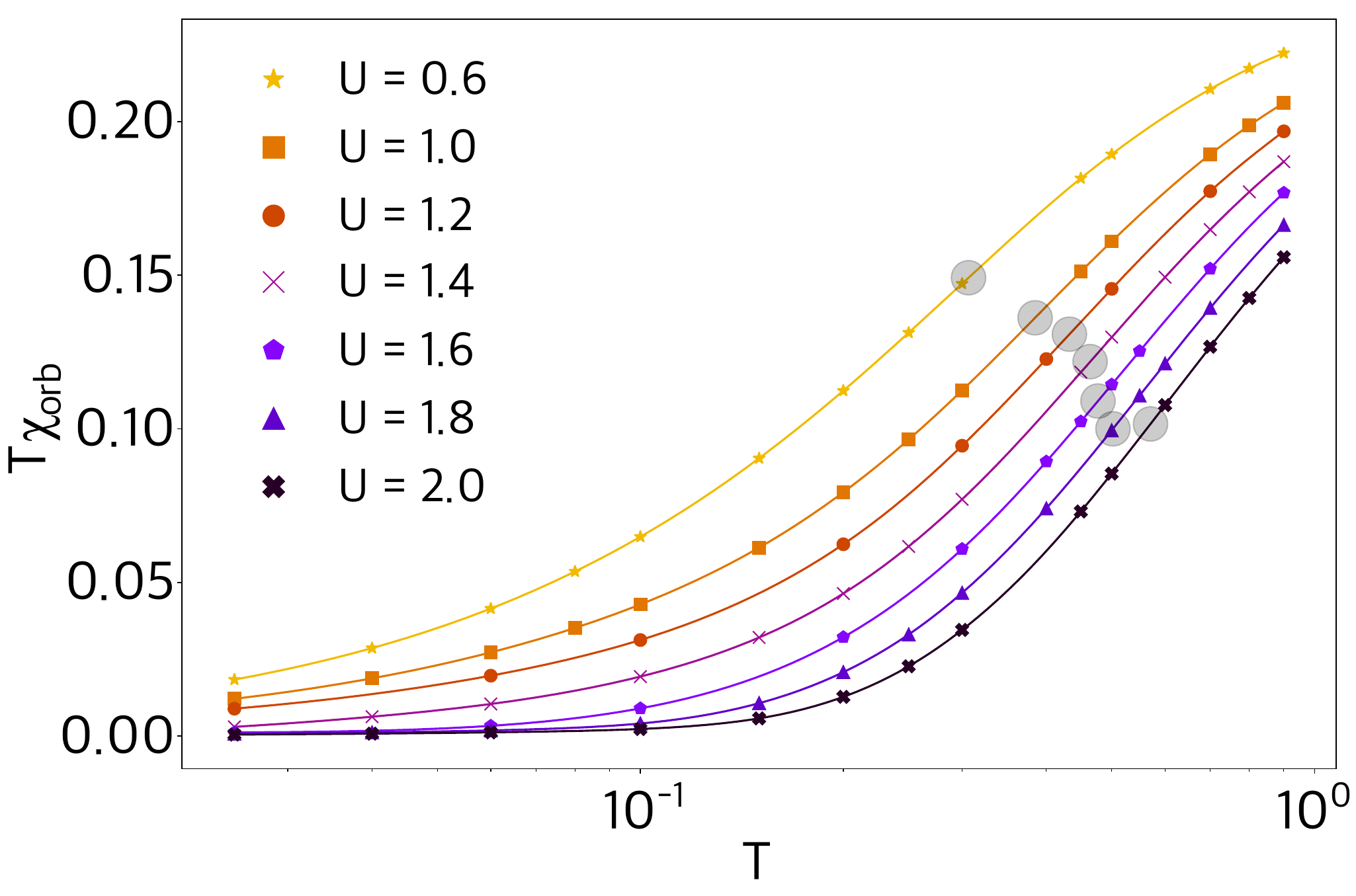}
\caption{(Color online). Temperature dependence of 
the square of the orbital moment for different values of the Coulomb interaction $U$. The Kondo temperature $T_{K,{\rm orb}}$ is obtained from the inflection points of the $T \chi_{\rm orb} (T)$, which are marked with gray circles.
}
\label{fig:chi_o}
\end{figure}

One of the hallmarks of Hund metals is the spin-orbital separation \cite{Hund3,Delft}, which implies strong difference of spin-and orbital Kondo temperatures. To obtain an orbital Kondo temperature $T_{\rm K,orb}$, which sets the upper boundary of the temperature of formation of local magnetic moments,
we consider the orbital susceptibility $\chi_{\rm orb}(T)$ as an indicator of changes in the interorbital balance of fillings (see Fig. \ref{fig:chi_o}). We identify $T_{\rm K,orb}$ with the inflection point of the logarithmic temperature dependence of the squared effective orbital momentum $\mu_{\rm orb}^2 = T\chi_{\rm orb}$.
One can see that the obtained temperature dependence implies suppression of the orbital susceptibility in the low temperature regime $T<T_{\rm K,orb}\sim 0.5t$, which will be mainly considered below.

\begin{figure}[b]
\includegraphics[width=1.0\linewidth]{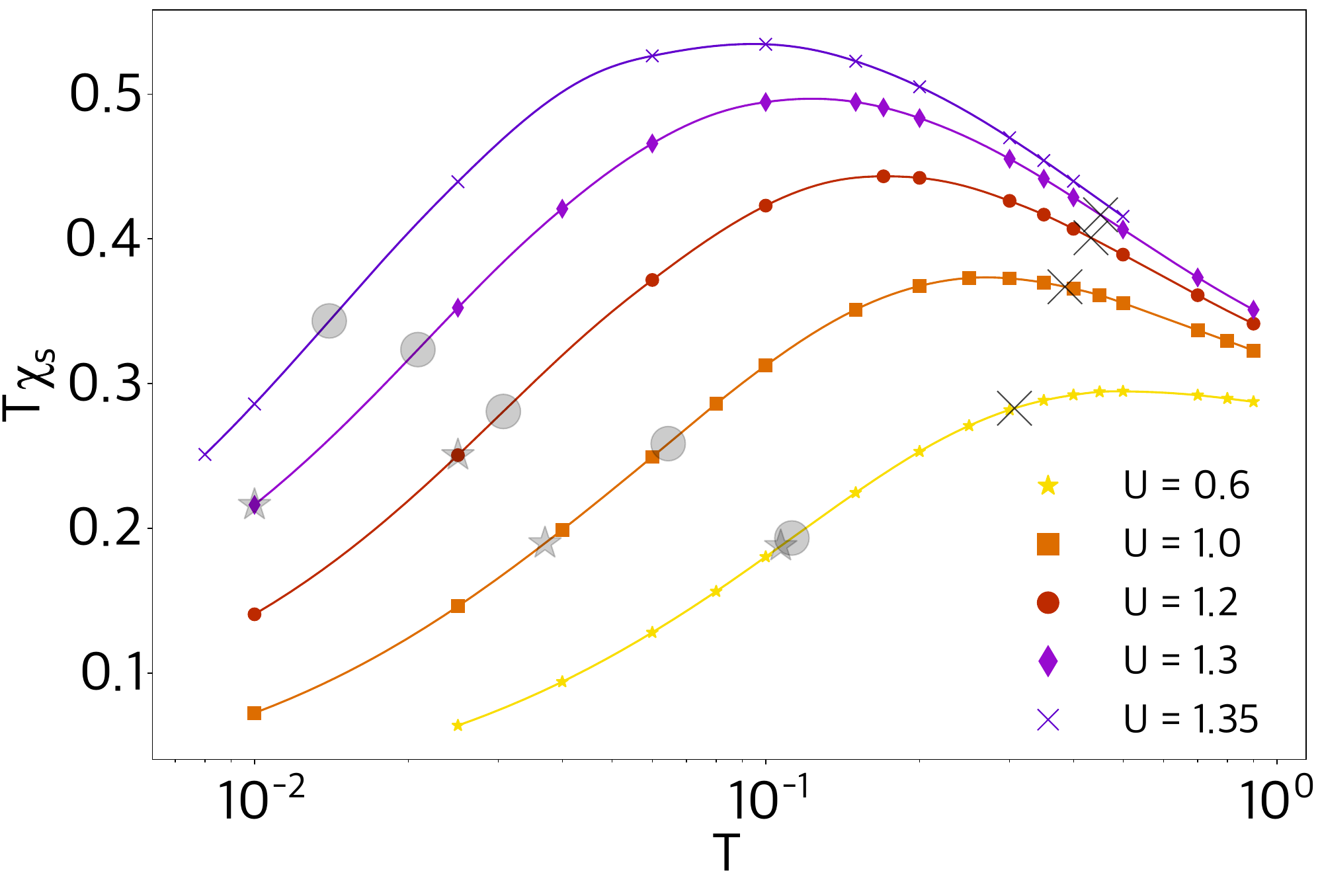}
\caption{(Color online). Temperature dependence of the square of the effective local magnetic moment $\mu^2_{eff} = T \chi_s$ for different values of the Coulomb interaction $U$. The Kondo temperatures $T_K$ obtained from the inflection points of the $T \chi_s(T)$ dependencies, are marked with gray circles. The orbital Kondo temperatures $T_{K,{\rm orb}}$ are marked by black crosses. Gray stars indicate the spin Kondo temperatures calculated with the fit of $T\chi_s(T)$ dependence to the $S=1$ Kondo model.
}
\label{fig:chi_s_kondo_s1}
\end{figure}

\begin{figure}[t]
\includegraphics[width=1.0\linewidth]{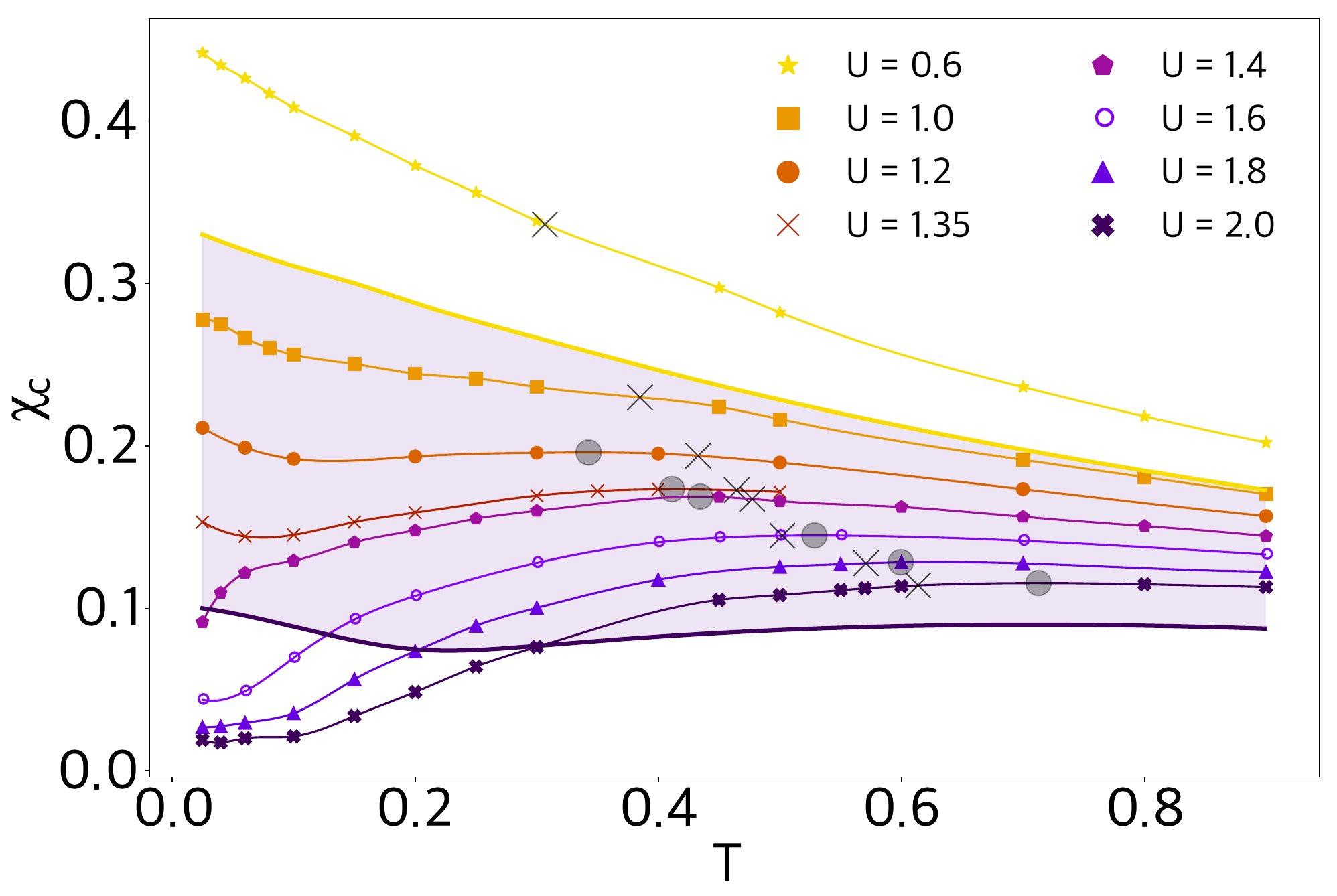}
\caption{(Color online). Temperature dependence of local static charge susceptibility $\chi_{\rm c}$ at various values of the Coulomb interaction $U$ and Hund coupling constant $J=U/4$. Gray circles indicate local maxima of $\chi_{\rm c}$. The orbital Kondo temperatures $T_{\rm K,orb}$ are shown by black crosses. The temperature dependencies of charge compressibility at $J=0$ and $U=0.6$, $U=2.0$ are shown by upper and lower solid lines without symbols. The shaded area marks the range of $\chi_{\rm c}$ values for interactions between $U = 0.6$ and $U = 2.0$. }
\label{fig:chi_c}
\end{figure}

On the other hand, the spin Kondo temperature can be determined from the temperature dependence of the effective local magnetic moment $\mu^2_{eff}(T) = T \chi_s(T)$, determined by the spin susceptibility.
Similarly to the orbital Kondo temperatures, we determine $T_K$ from the inflection point of the logarithmic temperature dependence $\mu^2=T \chi_s(T)$ (Fig. \ref{fig:chi_s_kondo_s1}), since the usually used fit to the universal dependence of $\mu(T)$ for the Kondo model (see, e.g., Refs. \cite{ToschiAnd1,Our1,Our2,Comment} for the single-band model), is not fully applicable to the considered two-band model in view of the spin anisotropy, induced by the density-density form of the interaction in Eq. (\ref{eq:hubbard}).
One can see that for not too small $U$ the corresponding Kondo temperatures are substantially smaller than $T_{\rm K,orb}$, but remain substantially higher
than in the single-orbital case \cite{Our1,Our2} because of the presence of more channels of screening.

To study the possibility of the formation and subsequent screening of local magnetic moments in the temperature range $T_K<T<T_{\rm K, orb}$, following previous works \cite{Our1,Our2}, we consider the temperature dependence of charge susceptibility 
(see Fig. \ref{fig:chi_c}).
One can see that for $U \geq 1.2$, a local maximum of $\chi_c(T)$ is observed, which characterizes the beginning of the formation of local magnetic moments in this region. With a further decrease in temperature, for $1.2 \leq U < 1.4$, a minimum  of $\chi_c(T)$ is observed with the subsequent increase, which is associated with the screening of local moments by conduction electrons, as was previously described for the single-band model \cite{Our1,Our2}. For $U \geq 1.4$, it is noticeable that the minimum of $\chi_c(T)$ is absent and the screening does not occur, which is the signature of an insulating-like behavior in this region. The orbital Kondo temperatures are also marked in the figure; as one can see, for not too small $U$ they are in close proximity to the beginning of the formation of local moments, which indicates a significant influence of interorbital correlations on the formation of this phase.

\begin{figure}[t]
\includegraphics[width=1.0\linewidth]{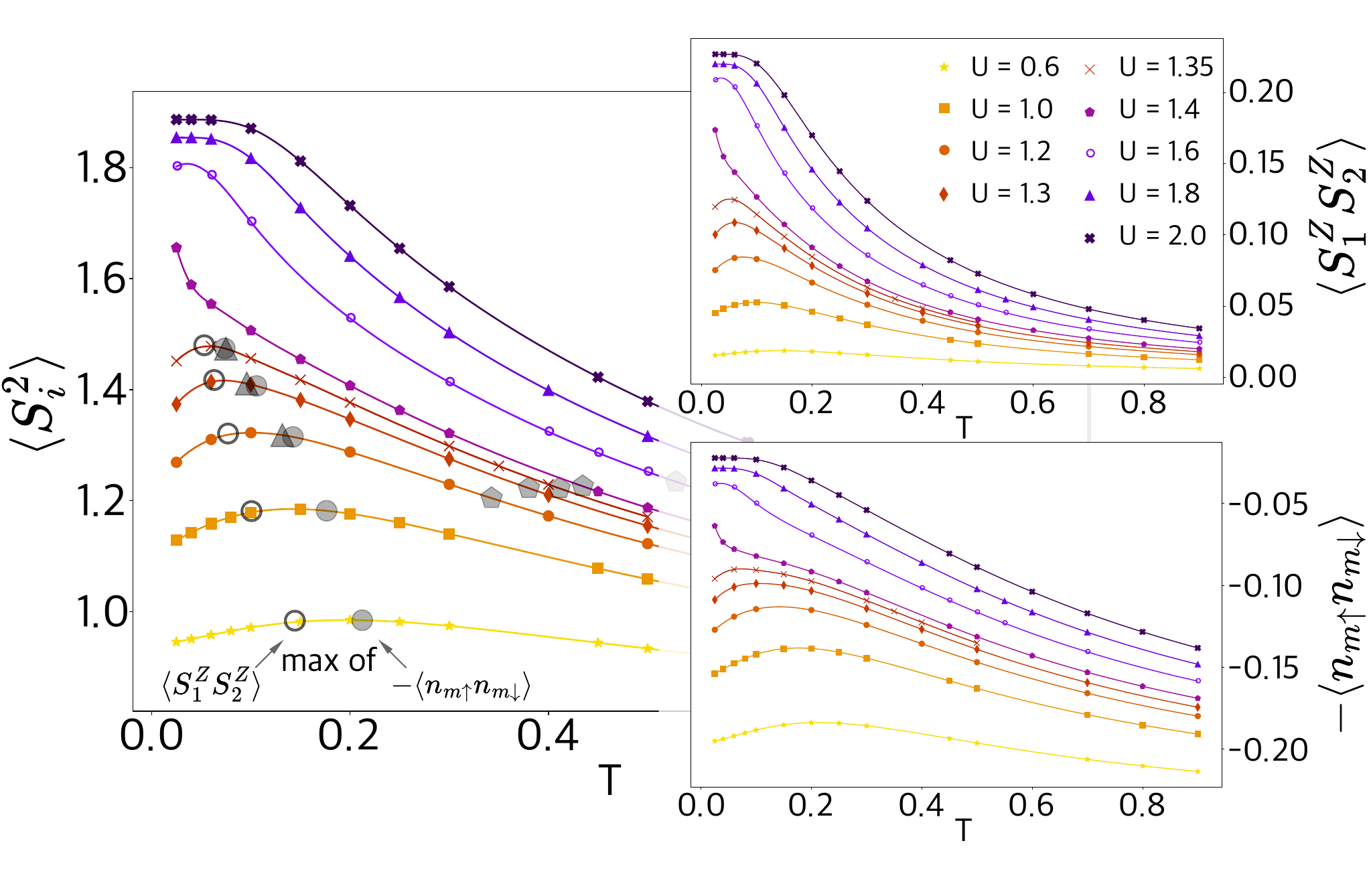}
\caption{(Color online). Temperature dependence of the square of local magnetic moment $\langle {\mathbf S}_i^2 \rangle$ (main plot), the interorbital spin correlator $\langle S^z_1 S^z_2\rangle$ (upper right plot), and the intraorbital (single-orbital) double occupation $- \langle n_{m \uparrow} n_{m \downarrow} \rangle$ (lower right) at various values of the Coulomb interaction $U$. Open and filled circles indicate maxima of $\langle S^z_1 S^z_2\rangle$ and $- \langle n_{m \uparrow} n_{m \downarrow} \rangle$ respectively. Filled triangles mark local minima of $\chi_{\rm c}$, and pentagons mark local maxima.}
\label{fig:s2}
\end{figure}

The obtained temperature dependencies of the orbital susceptibility can be compared to those with ``switched off" Hund exchange ($J=0$). In the latter case 
the local charge susceptibility predominantly increases with decreasing temperature, which indicates that the insulator region is located at higher values of $U$ ($U \sim 2.3$). Therefore, the noticable effect of Hund exchange is in the lowering of the metal-insulator Coulomb repulsion, which allows for having $T_{\rm K}\ll T_{\rm K,orb}$

To find the temperatures of (almost) full formation of local magnetic moments, we consider maxima of the mean square $\langle {\mathbf S}_i^2 \rangle$ (see Fig. \ref{fig:s2}, main plot). This allows us to distinguish  the regions of formation (increase of $\langle {\mathbf S}_i^2 \rangle$ with decrease of $T$), maximal saturation, and screening  of the LMM (decrease of $\langle {\mathbf S}_i^2 \rangle$). As one can see, in agreement with the observations from local charge susceptibility the screening occurs only when $U < 1.4$.

In contrast to the single-orbital case, there are two different contributions to $\langle {\mathbf S}_i^2 \rangle$ according to the Eq. (\ref{eq:s2}). The upper inset in Fig. \ref{fig:s2} shows the interorbital spin correlator $\langle S^z_1 S^z_2\rangle$, which originates from 
Hund exchange, while the lower inset shows the contribution of the temperature dependence of the single-orbital double occupation $- \langle n_{m \uparrow} n_{m \downarrow} \rangle$.
The latter quantity reflects the strength of the intra-orbital correlations. As one can see, the temperature dependencies of these quantities are similar, but their maxima are achieved at different temperatures (as noted on the main plot). That is, with decrease of temperature, correlations first set within each orbital, and then between the orbitals. One can also see that the minima of the temperature dependence of charge susceptibility, discussed above, correspond to the maxima of the intraorbital correlations.
While in the single-band model the maxima of the intra-orbital correlations (which also correspond to the maxima of charge compressibility) were associated with the onset of the screening, presence of the interorbital contribution in the two-band model shifts the actual onset of the screening to lower temperatures. We therefore further use the maximum of the full squared moment $\langle {\mathbf S}_i^2 \rangle$ rather than the minimum of charge compressibility as a criterion of the onset of the screening. This criterion also appears applicable in the doped case, in contrast to the charge compressibility, which is sensitive to the presence of free charge carriers \cite{Our2}, see further discussion below in Sect. IV. 

\begin{figure}[t]
\includegraphics[width=1.0\linewidth]{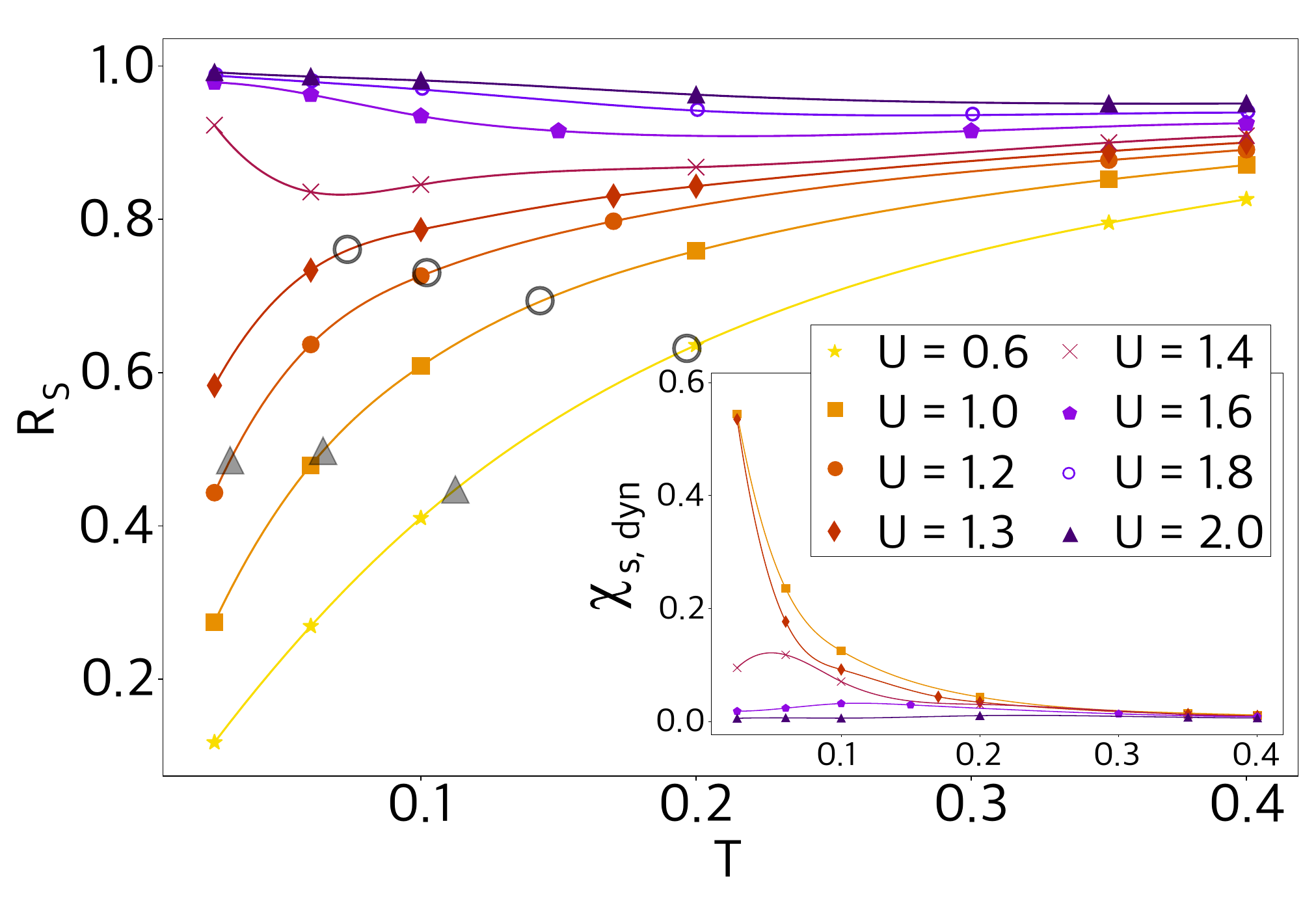}
\caption{(Color online). Main plot: temperature dependence of ``frozen spin ratio" $R_s$, Eq. (\ref{Froz}) for different values of the Coulomb interaction $U$. Open circles indicate maxima of $\langle {\mathbf S}_i^2 \rangle$ and filled triangles mark spin Kondo temperatures $T_K$. Inset: temperature dependence of dynamic contribution to local spin susceptibility $\chi_{s, dyn}$}
\label{fig:r_s}
\end{figure}

\begin{figure*}[t]
\includegraphics[width=0.8\linewidth]{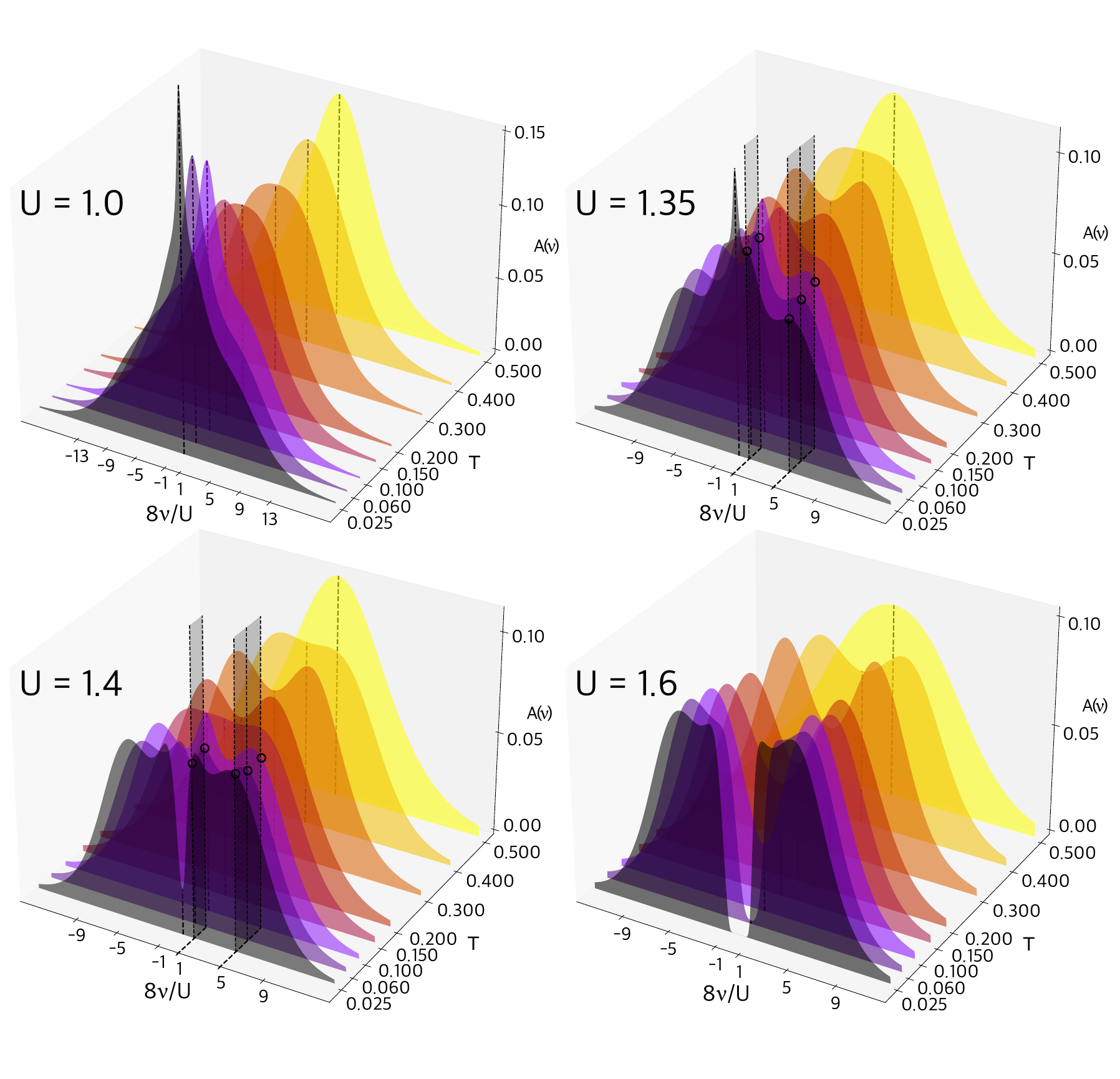}
\caption{(Color online). Temperature evolution of local spectral functions $A(\nu)$ at different $U$. The dotted lines mark the $\nu=0$ boundary, indicating the density of electronic states at the Fermi level. {The vertical planes show
characteristic excitation energies corresponding to atomic states $|\sigma,\sigma \rangle$ ($5U/8$) and $\ket{\uparrow \downarrow, 0}$ ($U/8$).}}
\label{fig:spectral}
\end{figure*}

Let us compare the obtained temperature dependencies of charge- and spin susceptibilities to those of the frozen spin ratio (\ref{Froz}), see Fig. \ref{fig:r_s}. At the transition from the metallic-like to insulating-like state, obtained above from the temperature dependence of charge susceptibility, the temperature dependence of frozen spin ratio also changes from decreasing to increasing with the decrease of temperature at low $T$, which corresponds to difference between screened and unscreened LMM in the metallic-like and insulating-like regimes. The value $R_s>1/2$ was suggested in pioneering study of Ref. \cite{Freezing} as a criterion for the local magnetic moment formation. One can see that this criterion is fulfilled in the insulating-like regime and in the metallic regime it is fulfilled above {the temperature of full screening of LMM (i.e. the Kondo temperature $T_K$). At the temperature of the beginning of screening, which was associated above with the maximum of the square of instantaneous of local magnetic moment $\langle {\mathbf S}_i^2 \rangle$, we observe  a rapid decrease in $R_s$. Therefore, with the onset of screening both the amplitude of the total moment and its stability over time $\tau$ decreases. This also confirms the consistency of the methods for determining the behavior of the LMM. At high temperatures we observe a plateau $R_s\sim 0.9$, which appears due to consideration of a short imaginary time interval $\beta / 2$ at large $T$.} Therefore, the temperature dependence of the frozen spin ratio does not reflect the onset of local magnetic moment formation. 

{The stability of local magnetic moments can be also determined from the dynamic contribution to the spin correlation function $\chi_{s, {\rm dyn}} = \chi_s - \beta \left\langle S^z(\beta/2) S^z(0) \right\rangle$, discussed earlier in Ref. \cite{WernerAndHoshino3}. As one can see from the inset of Fig. \ref{fig:r_s}, small values of dynamic part of the susceptibility are observed in the unscreened LMM regime. On the contrary, as the interaction decreases, the dynamic contribution increases in the screened LMM regime, which is also in agreement with the $R_s$ dependence discussed above.}

\begin{figure}[t]
\includegraphics[width=0.90\linewidth]{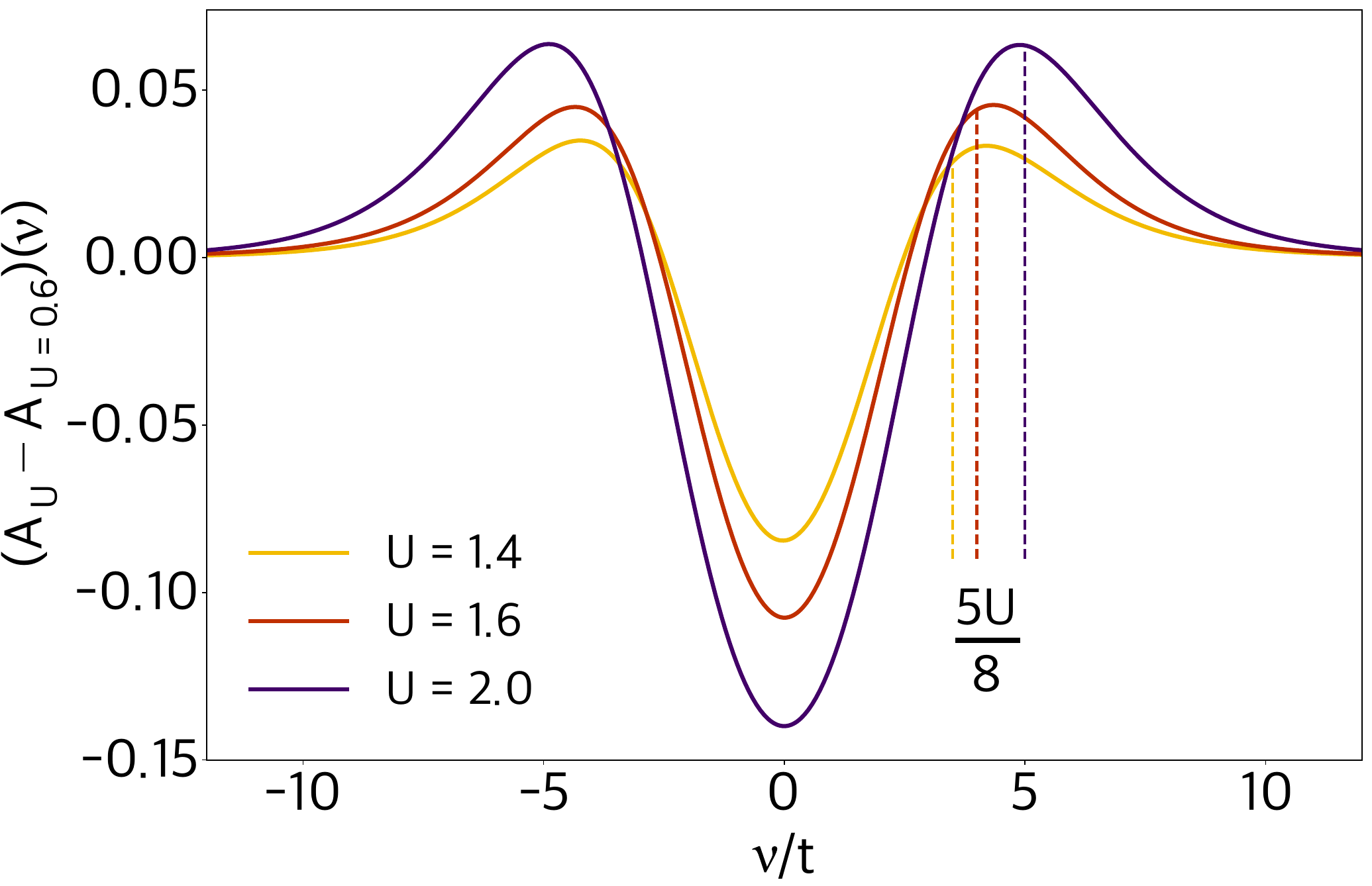}
\caption{(Color online). Frequency dependence of the difference between spectral functions $A_U - A_{U = 0.6}$. The dotted lines mark the boundaries of the excitation energy $(U+J)/2 = 5U/8$ (for each $U$), corresponding to the state $|\sigma,\sigma \rangle$.}
\label{fig:spectral_diff}
\end{figure}

In order to 
see peculiarities of the electronic properties in various regimes, we also consider the frequency dependencies of the local spectral functions, which represent the density of states of interacting electrons (see Fig. \ref{fig:spectral}). At small $U=1.0$ we find van Hove singularity of the density of states at low temperatures which is smeared by thermal fluctuations, so that the corresponding evolution of the density of states is only weakly affected by the interaction effects. However, at somewhat larger $U=1.35$ we find formation of Hubbard subbands at sufficiently low temperatures. {To analyze in more detail the shape of the spectral functions} in the temperature range $0.2<T<0.4$ we show in Fig. \ref{fig:spectral_diff} the difference of spectral function $A(\nu)$ at $T = 0.3$ to that at small $U=0.6$ (we have verified that further decrease of the interaction does not affect the result). One can see that the interaction $U$ leads to redistribution of the spectral weight from the center of the band (i.e. from the Fermi level) to the energy $\simeq\pm(U+J)/2 = \pm 5U/8$ (at the corresponding $U$). As we show in Appendix, this excitation energy refers to
the energy of adding (removing) electrons to (from) the state $|\sigma,\sigma \rangle$, showing that in this region, the formation of local magnetic moments actually occurs. 

\begin{figure}[b]
\includegraphics[width=1.0\linewidth]{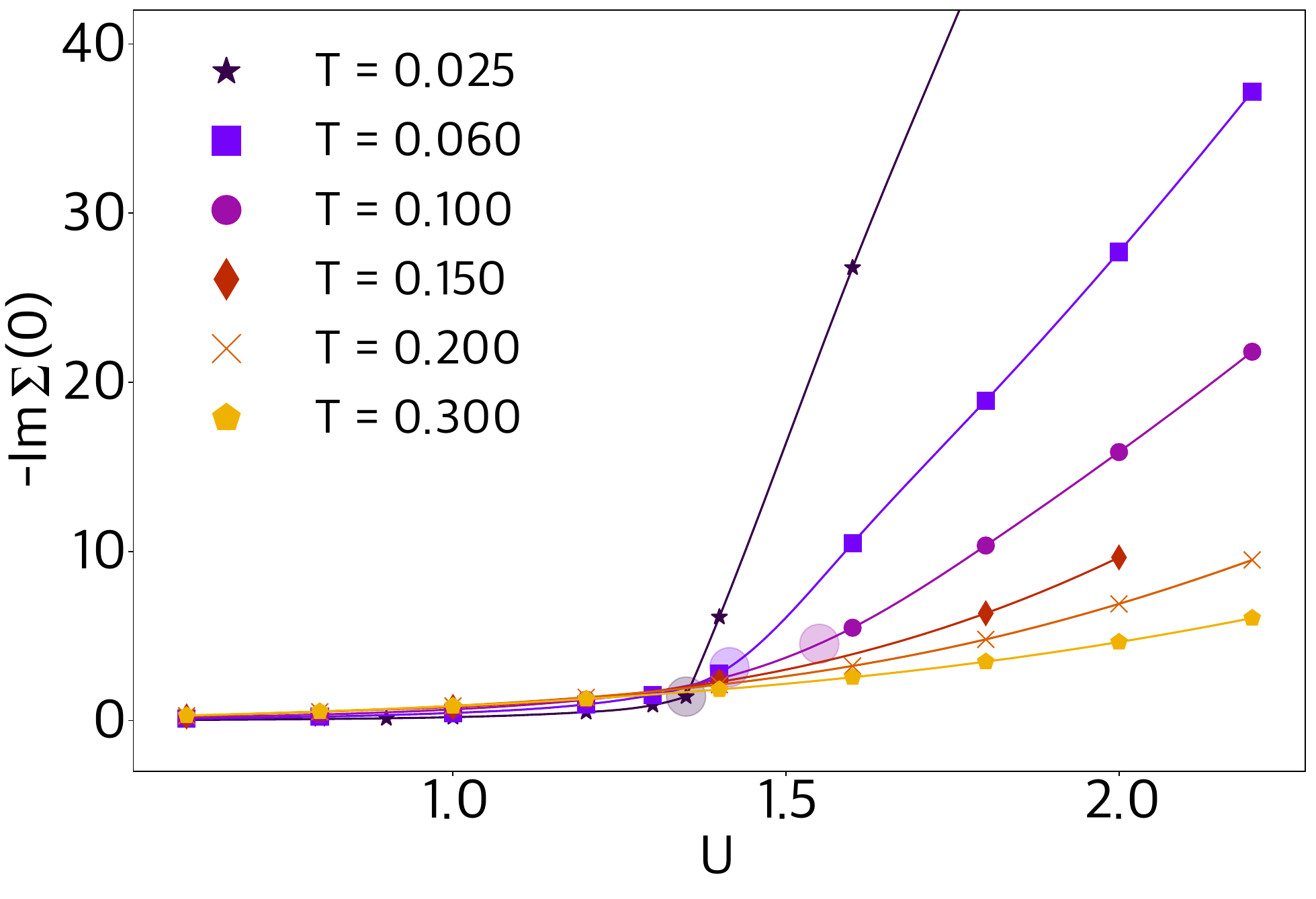}
\caption{(Color online). Interaction dependence of the imaginary part of zero-frequency self-energy at various values of the temperature $T$. Colored circles indicate points of rapid growth of $\Gamma=-{\rm Im}\Sigma(0)$, obtained from largest curvature.}
\label{fig:im_sigma0}
\end{figure}

{At lower temperatures in the metallic region (Fig. \ref{fig:spectral}),
both, the quasiparticle peak and the Hubbard subbands at the energy $\pm(U+J)/2$ exist.}
Their simultaneous appearance shows  screening of local magnetic moment, cf. Refs. \cite{Our1,Our2}. 
In contrast to the single-orbital model, the 
narrower side peaks with the energies $\pm (5J-U)/2=\pm U/8$ appear as well in this regime. The latter peaks correspond to removing (or adding) electron from (to) the $\ket{\uparrow \downarrow, 0}$ state, which appear as a part of the processes of local magnetic moment screening.
On the other hand, at large $U$ we find the gap of the spectral function, which is continuously filled with temperature.

\begin{figure}[t]
\includegraphics[width=1.0\linewidth]{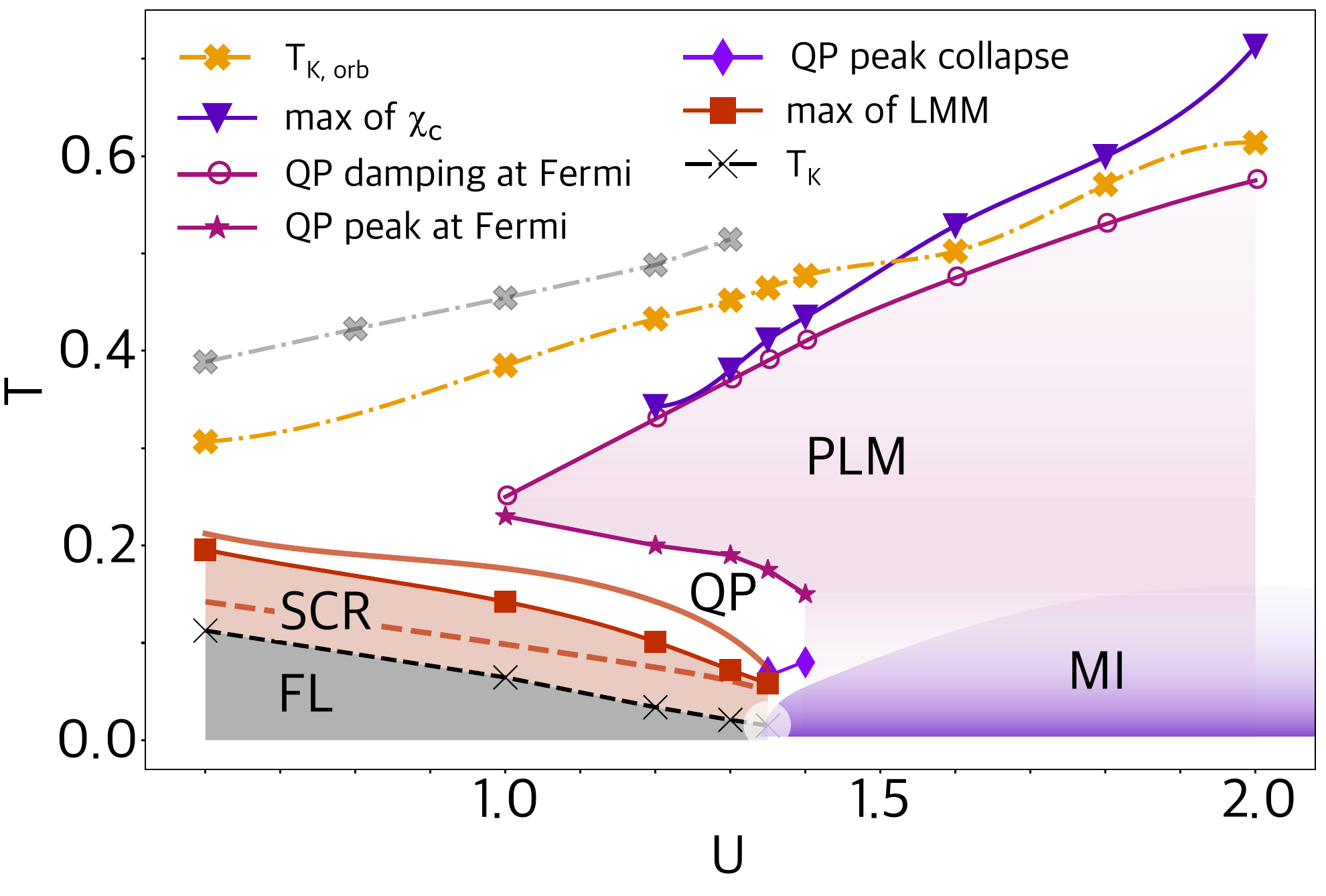}
\caption{(Color online). The obtained phase diagram at half filling. The orange dash-dotted line (crosses) shows the orbital Kondo temperatures $T_{K,{\rm orb}}$ {(grey dash-dotted line (crosses) corresponds to $T_{K,{\rm orb}}^\omega/2$)}. The local maxima of $\chi_c(T)$ are shown by violet line (triangles). The preformed local moment (PLM) region, bounded by the increase of QP damping (open circles) and appearance of QP peak of the spectral function (stars), is indicated by the purple shaded area. The violet line (diamonds) marks the suppression of QP peak. The red shaded region (SCR) corresponds to the LMM screening regime, bounded by the maxima of the instantaneous magnetic moment (squares). The solid red line (without squares) indicates the maxima of the intraorbital moment, and the dashed line indicates the maxima of interorbital correlations. The dark shaded area (FL) indicates the Fermi liquid phase, bounded by the respective Kondo temperatures (crosses). The Mott insulator (MI) region is shown by the violet area determined from the quasiparticle damping. Between the PLM and SCR regions there is a regime of appearance of quasiparticle peak in the absence of pronounced screening (QP).}
\label{fig:phase_diagram_hund}
\end{figure}

The most complex behavior is observed near the Mott transition (see, e.g., the result for $U = 1.4$). It can be seen that at low temperatures there is a strong dip at the Fermi level due to insulating behavior, which, however, turns into a quasiparticle peak with increasing temperature. 
Next, a region of the formed local moment is observed, which is also noticeable from the suppression of the spectral function at the Fermi level. 

The transition to the insulating-like state 
at $U \approx 1.4$ 
can be further confirmed by loss of coherence of the quasiparticles
as reflected by the quasiparticle damping, estimated as an imaginary part of the electronic self-energy $\Gamma=-{\rm Im}\Sigma(0)$, {which is obtained from linear extrapolation of the self-energy at lowest Matsubara frequencies to zero frequency, see Fig. \ref{fig:im_sigma0}.}
Indeed, crossing the border $U \approx 1.4$, the damping rapidly increases, while to the left of this border, as can be seen, $\Gamma$ changes only slightly with a change in $U$ and $T$. We associate the transition to the insulating-like behavior with the increase of damping, characterised by the maximum of the second derivative. One can see that the obtained boundary between the metallic-like and insulating-like behavior coincides with that, obtained from the change of the temperature dependence of local charge compressibility, which is suppressed at $U\gtrsim 1.4$.



The obtained boundaries of various regimes are shown in the phase diagram for half filling in Fig. \ref{fig:phase_diagram_hund}. As in earlier works \cite{Our1,Our2}, at sufficiently large temperatures and Coulomb interactions we find the preformed local moment regime (PLM). The upper temperature boundary of PLM regime, determined 
by the sharp increase of the quasiparticle damping, 
is quite close to the line of charge compressibility maxima and, at sufficiently large $U$, to the orbital Kondo temperature $T_{\rm K, orb}$, which indicates the correctness of determining the characteristic temperatures $T_{\rm K, orb}$. This fact also confirms considering the orbital screening temperature $T_{\rm K, orb}$ as start of the local magnetic moment formation in the presence of Hund exchange.
{We have verified that the orbital Kondo temperatures $T^\omega_{\rm K,orb}$, obtained from the maximum of the frequency dependence of the spectral weight ${\rm Im}\chi_{\rm orb}(T,\omega)$ (cf. Ref. \cite{Hund3}, see Appendix \ref{TKorb1} for more details) possess approximately the same $U$-dependence, as $T_{\rm K,orb}$, being related to the Kondo temperatures from the static susceptibility by $T^\omega_{\rm K,orb}\simeq 2.4T_{\rm K,orb}$. 
Both methods 
yield approximately linear increase of the orbital Kondo temperature with $J$, $T_{K,{
\rm orb}}\simeq J+0.13$ ($J=U/4$ in this study).}
It
{reflects the breakdown of LMM, formed by electrons persisting in two different orbitals coupled by Hund exchange interaction, by thermal fluctuations at $T>T_{K,{\rm orb}}$.}

\begin{figure}[b]
\includegraphics[width=1.0\linewidth]{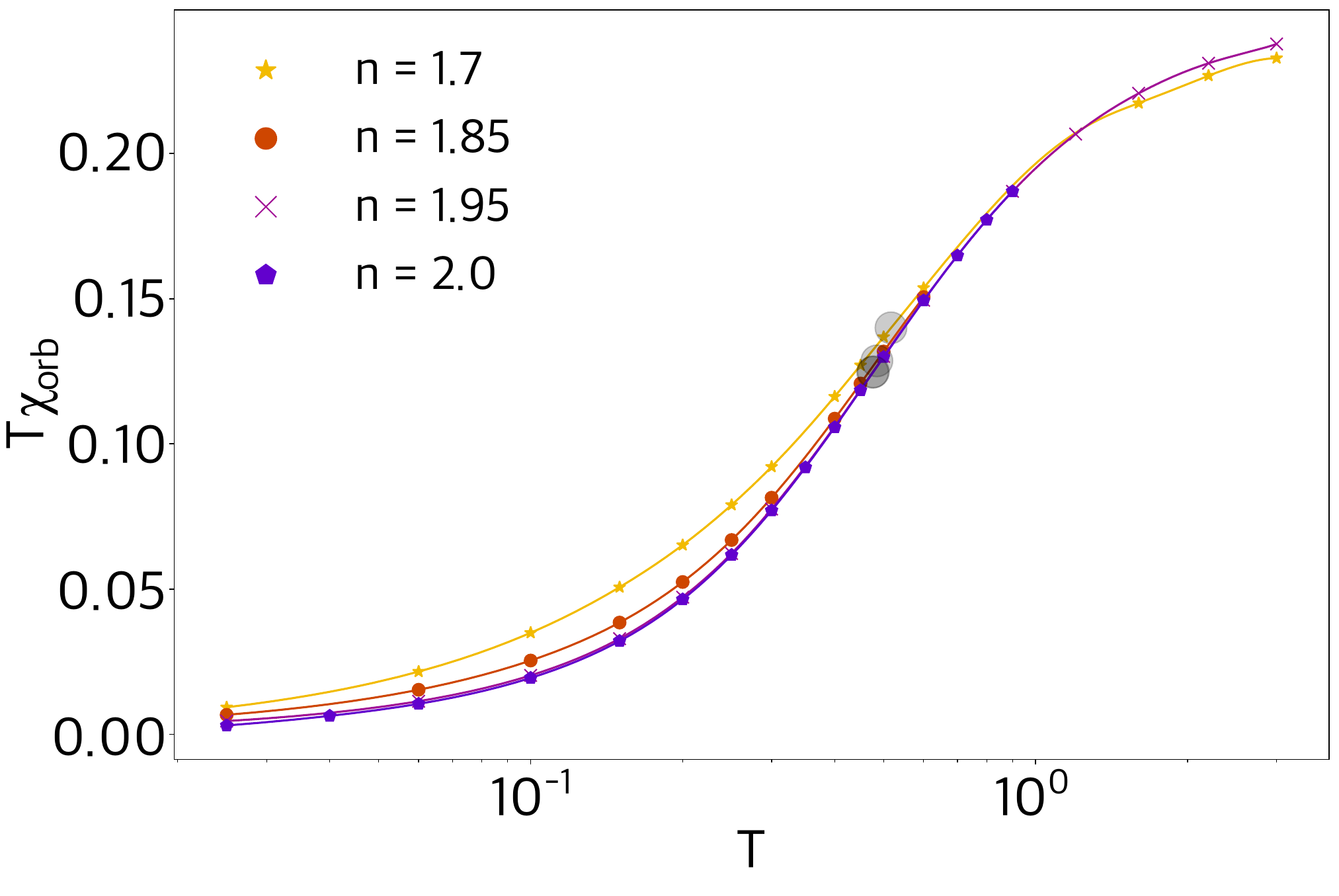}
\caption{(Color online).  Temperature dependence of the square of the orbital moment for various fillings $n$ ($U=1.4$). The Kondo temperature $T_{K,{\rm orb}}$ is obtained from the inflection points of the $T \chi_{orb} (T)$, which are marked with filled circles.}
\label{fig:chi_o_n}
\end{figure}

With decrease of temperature PLM regime changes first to the quasiparticle regime (QP), which  precedes the screening of LMM (SCR regime) similarly to the single orbital case \cite{Our2}. However, for considered two-band model this regime ``penetrates" into the region above the insulating phase and has the boundary, which behaves non-monotonically with change of temperature and Coulomb repulsion. {This non-monotonicity is likely related to the first-order ground state Mott transition, discussed previously in Ref. \onlinecite{MediciTwoBand}}. At sufficiently strong interaction (above the mott insulating phase) the QP peak collapses with decreasing temperature. Finally, in the FL regime the fully screened local moment Fermi liquid state occurs.

\section{Results away from half filling}

\begin{figure}[t]
\includegraphics[width=1.0\linewidth]{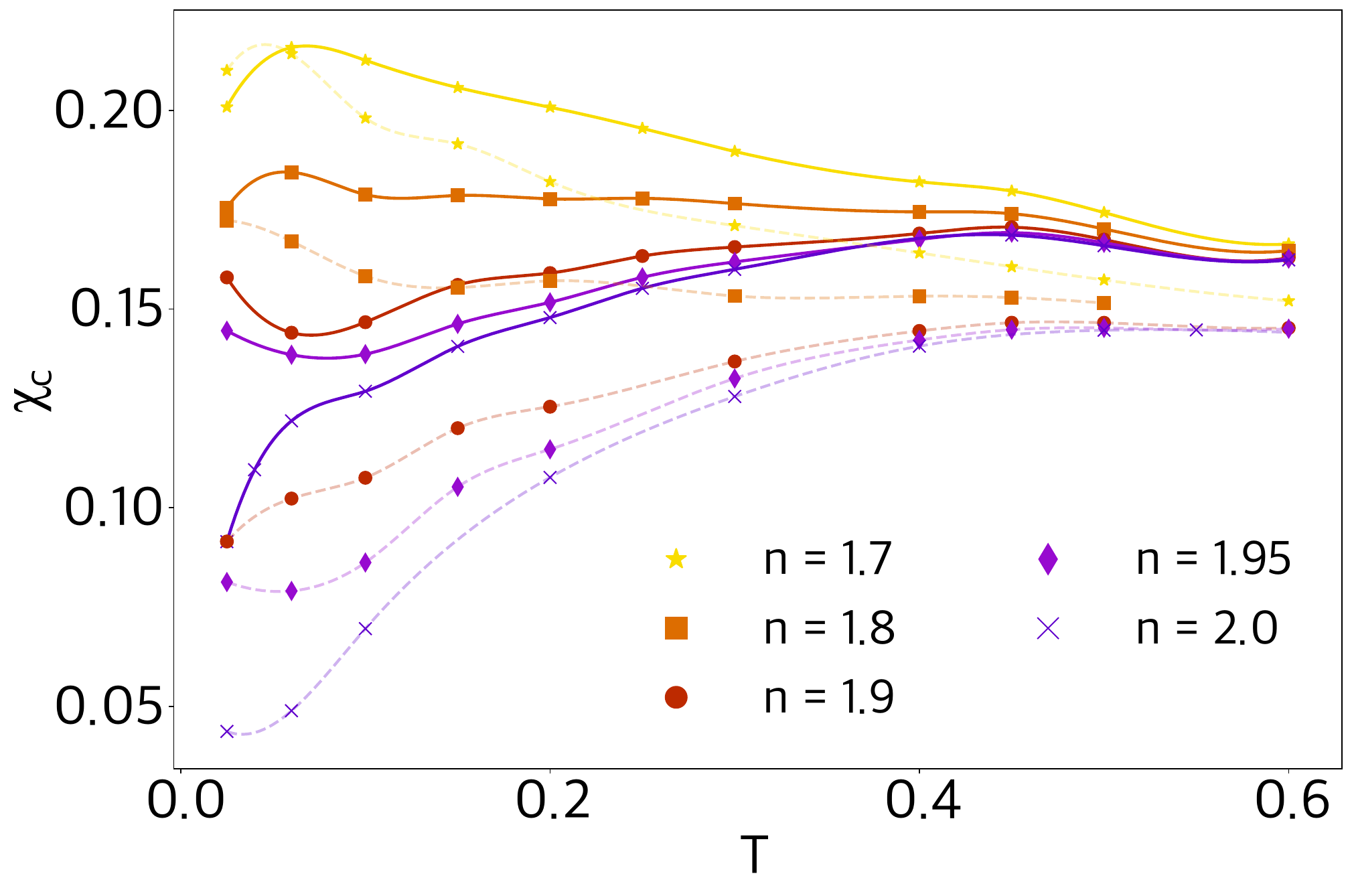}
\caption{(Color online). Temperature dependence of local static charge susceptibility $\chi_c$ for various fillings $n$ (solid lines correspond to $U = 1.4$ and dotted lines correspond to $U = 1.6$).}
\label{fig:chi_c_n}
\end{figure}

\begin{figure}[b]
\includegraphics[width=1.0\linewidth]{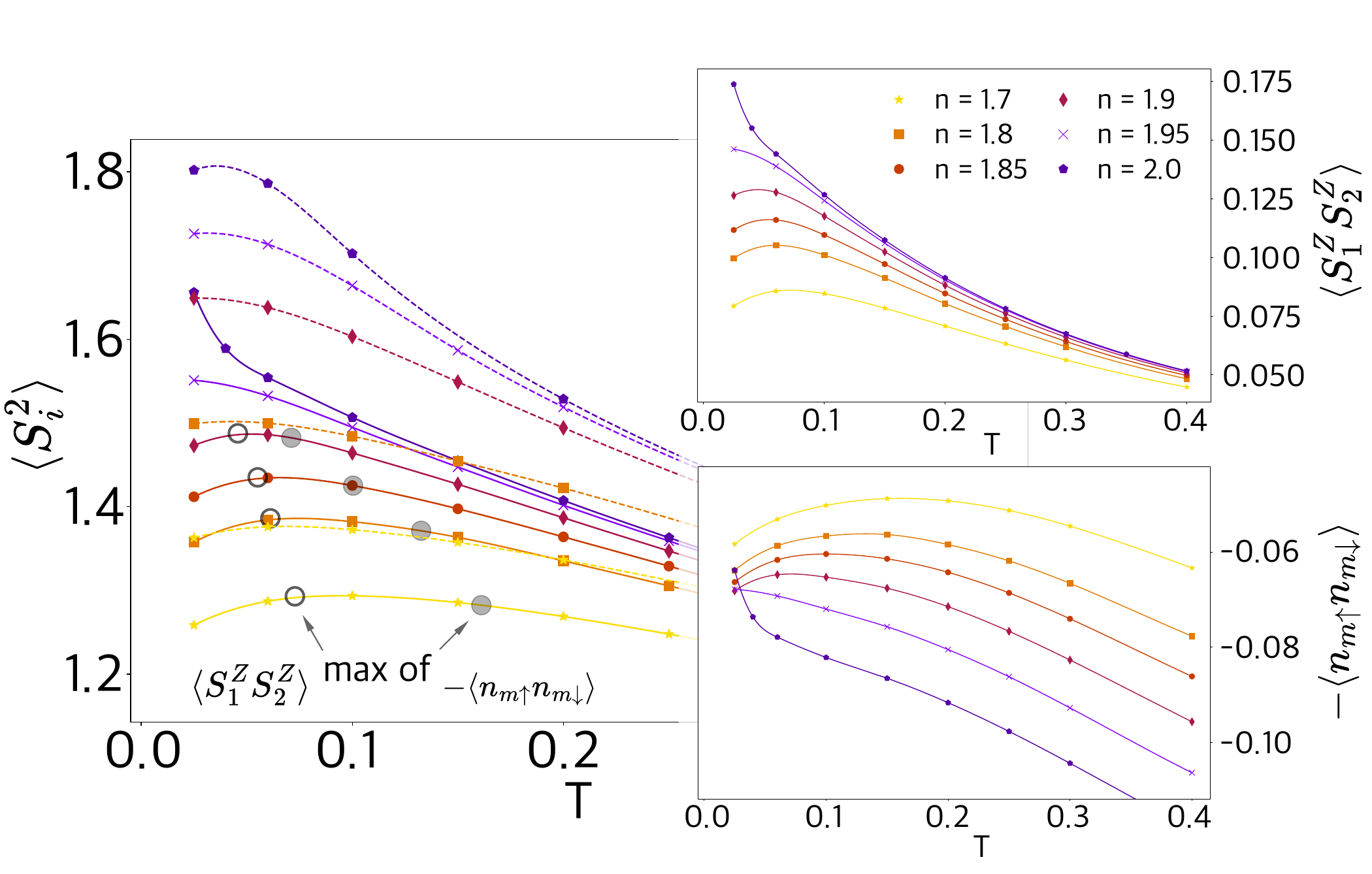}
\caption{(Color online). Main plot: temperature dependence of the square of local magnetic moment $\langle {\mathbf S}_i^2 \rangle$ for $U=1.4$ (solid lines) and $U=1.6$ (dashed lines). Insets: interorbital spin correlator $\langle S^z_1 S^z_2\rangle$ (top) and intraorbital (single-orbital) double occupation $- \langle n_{m \uparrow} n_{m \downarrow} \rangle$ (bottom) at various fillings $n$ for $U = 1.4$. Open and filled circles indicate maxima of $\langle S^z_1 S^z_2\rangle$ and $- \langle n_{m \uparrow} n_{m \downarrow} \rangle$ respectively.}
\label{fig:s2_n}
\end{figure}

\begin{figure*}[t]
\includegraphics[width=0.85\linewidth]{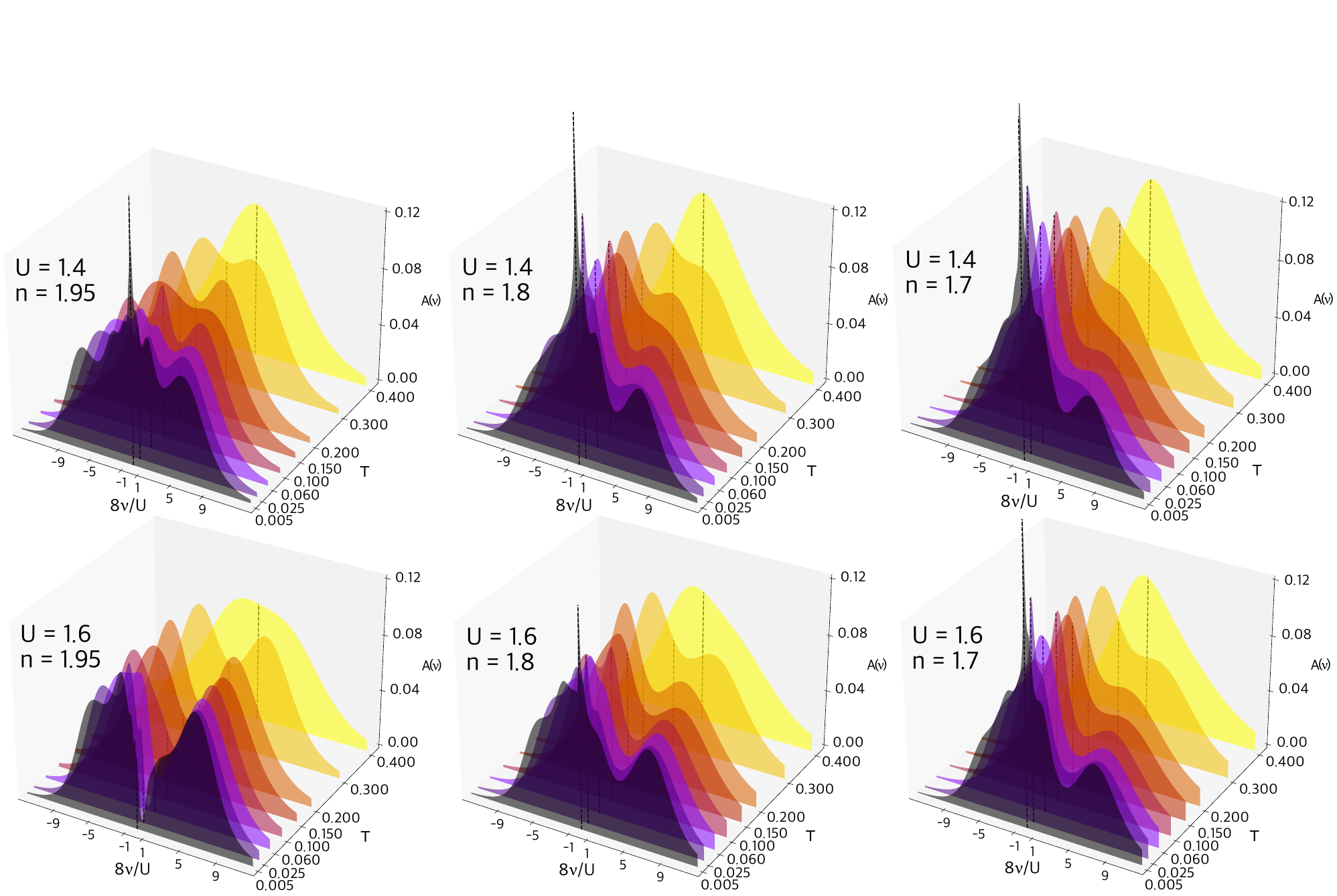}
\caption{(Color online). Temperature evolution of spectral functions $A(\nu)$ at different interaction $U$ and concentration $n$. The dotted lines mark the $\nu=0$ boundary, indicating the density of electronic states at the Fermi level.}
\label{fig:spectral_n}
\end{figure*}

In this section, we consider what changes in the above discussed observables with doping. In accordance with the consideration of previous Section, we determine first the orbital Kondo temperatures from the inflection points of the $T \chi_{\rm orb} (T)$ dependence (Fig. \ref{fig:chi_o_n}). It is noticeable that in agreement with earlier results \cite{Hund3}, $T_{\rm K, orb}$ is almost unchanged with doping. Therefore, orbital correlations are not strongly dependent on the concentration of electrons.

We consider in the following the case in the  vicinity of the metal-insulator transition ($U = 1.4$) and the insulating phase at half filling ($U = 1.6$). The temperature dependence of charge compressibility (Fig. \ref{fig:chi_c_n}) changes with doping similarly to the single-band case \cite{Our2}. In particular, instead of the maximum (start of the formation of LMM), followed by the minimum, corresponding to beginning of screening of LMM, for $n\lesssim 1.8$ it increases with decrease of temperature due to sufficiently large number of free charge carriers (i.e. holes), and has a maximum at approximately the same temperature, where the minimum of the susceptibility was observed for larger fillings. This change of temperature dependence was identified in Ref. \cite{Our2} with the coherent hole motion (CHM) regime, which is present in the two-orbital model as well.   


The temperature dependence of the average square of the local magnetic moment  $\langle {\mathbf S}_i^2 \rangle$ is similar to half filling (Fig. \ref{fig:s2_n}, main plot). In particular, near half filling it monotonously increases with decreasing temperature, which shows the formation of the LMM. With increase of doping the maximum of the temperature dependence of local magnetic moment is formed, showing onset of screening. In stark contrast to the single-orbital case \cite{Our2}, the temperature of the maximum of LMM formation only weakly changes with doping. Resolving the intra- and interorbital contributions (Fig. \ref{fig:s2_n}, insets) shows that the maximum of the intraorbital contribution $-\left\langle n_{im\uparrow} n_{im \downarrow} \right\rangle$ strongly shifts to larger temperatures similarly to the single-orbital case, but in the two-orbital model it is compensated by the interorbital contribution $\langle S_{i,1}^z S_{i,2}^z\rangle$, which maximum remains unchanged with doping. Therefore, the interorbital correlations, originating from Hund exchange, provide a decisive role in preserving LMM with doping. 


The evolution of spectral functions in the doped regime is of particular interest for determining presence of LMM. In the single-band case it was found \cite{Our2} that local magnetic moments at finite doping are present only in screened form, which corresponds to having narrow quasiparticle peak at the Fermi level at low temperatures. However, as it is discussed above in this section, in the two-orbital case the effect of Hund interaction is significant for providing stability of LMM. The spectral functions $A(\nu)$ are shown in Fig. \ref{fig:spectral_n} for both $U=1.4$ and $U=1.6$. At small doping and temperatures $T \sim 0.2$ -- $0.3$ we find a dip of the spectral functions at (or close to the) the Fermi level, which corresponds to the LMM formation. Thus, despite the delocalization of charge carriers with decreasing filling, the Hund exchange between orbitals preserves the LMM. Qualitatively, this can be explained by the necessity to increase the energy of electrons by Hund exchange when breaking the LMM, which is comparable to the hopping parameter. Therefore, in contrast to the single-band model, in multi-band model the hopping becomes much less effective for breaking the LMM. This is an essence of Hund metal behavior. 

\begin{figure}[t]
\includegraphics[width=1.0\linewidth]{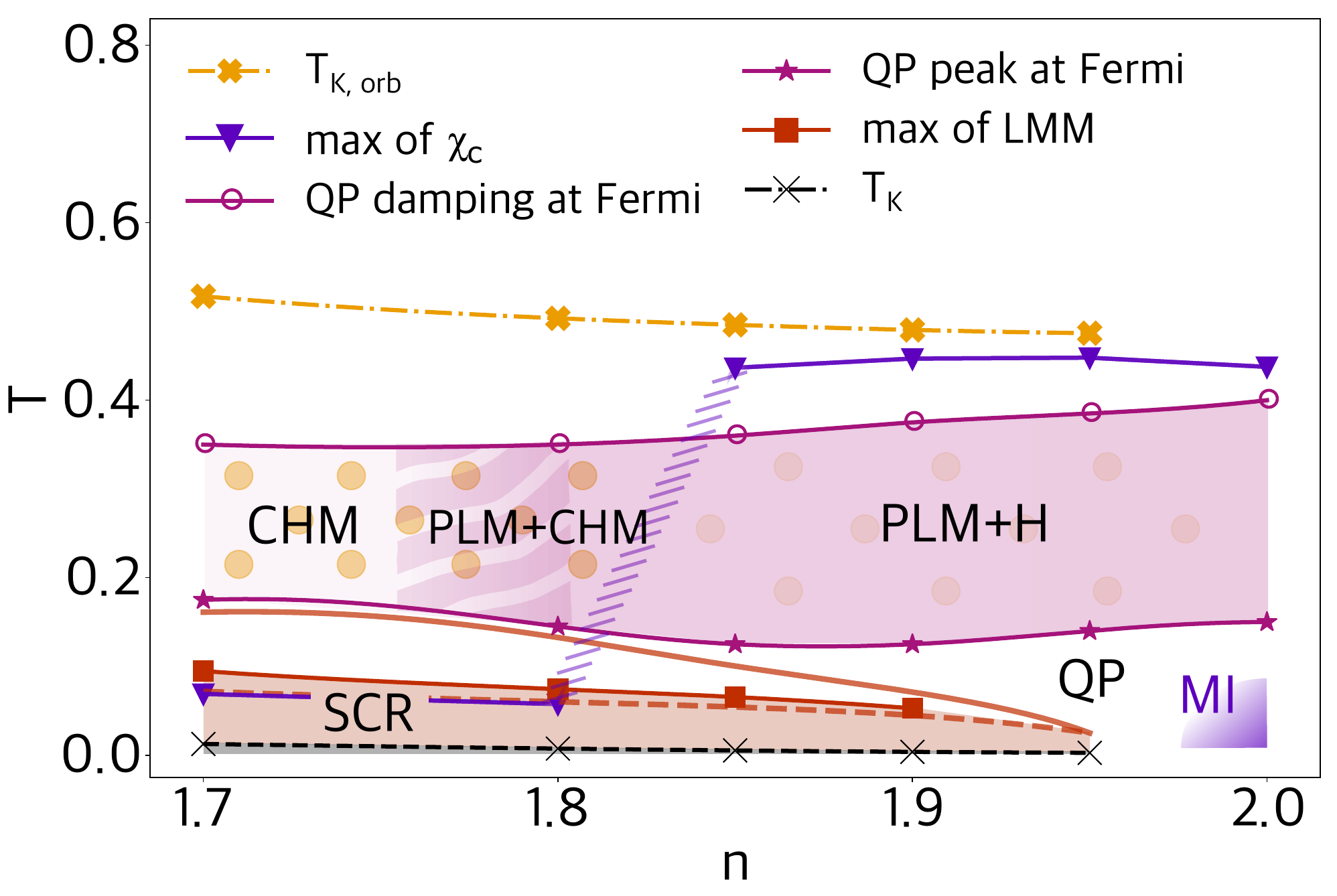}
\vspace{-0.75cm}
\caption{
(Color online). The phase diagram away from half filling for $U=1.4$. The symbols on the diagram are the same as in the half filled case in Fig. \ref{fig:phase_diagram_hund}. 
The PLM+H (PLM+CHM) phases correspond to the preformed local moments, accompanied by localized (moving) holes. 
The Mott insulator (MI) region is shown by the violet area. Between the PLM and SCR modes there is a regime of appearance of quasiparticles before screening (QP).
}
\label{fig:diagram_n}
\end{figure}

To summarize the results for the doped system, we consider the phase diagram in Fig. \ref{fig:diagram_n}. We consider the case near the transition ($U=1.4$), for $U=1.6$ the phase diagram looks similar except for the extent of the insulator and screening regions. As it was noted above, the orbital Kondo temperature remains almost unchanged with doping (cf. Ref. \cite{Hund3}) and only increases slightly, which implies an increase of the distribution of electrons in different orbitals (see Eq. (\ref{eq:chi_orb})). At small doping the temperatures of charge compressibility maxima remain close to $T_{K, orb}$, as in the undoped case. With further decrease of temperature, a wide region of the formation of a local magnetic moment (PLM) is observed, obtained from observations of electronic spectral functions. Despite presence of holes, at small doping we find sufficiently wide PLM+H regime, where owing to the Hund exchange, delocalized charges do not destroy completely the LMM, which coexists with free charge carriers (i.e. holes). With significant doping ($n<1.85$), similarly to the single-orbital case, the maximum of temperature dependence of charge compressibility is changed by its minimum. The corresponding region $n<1.85$, which previously was attributed to the coherent hole motion (CHM) regime, contains however a part with preserved local magnetic moments (PLM+CHM regime), 
{which is characterized by the suppressed spectral weight at the Fermi level and reflects the presence of well-formed local moments with almost freely moving charge carriers.} 
With further increase of doping, the spectral weight at the Fermi level continuously increases, which  
characterizes the destruction of local moments, which marks the onset of CHM regime.
At low temperatures a screening region appears. We define the boundary of the screening region by reaching the maximal value of the LMM, and, as can be seen, the maximum of the intra-orbital correlations is reached at much larger temperatures, while the effect of interorbital correlation persists shifts the maximum to much lower temperatures. Importantly, the temperature of the onset of screening is suppressed in comparison to the single-orbital model \cite{Our2} and separated from the region of formation of LMM by narrow QP region, which is present in the single-orbital model only at half filling \cite{Our2}.
Fully screened local moments are represented by a Fermi liquid below the Kondo temperature $T_K$.




\section{Conclusions}
We have considered the effect of local magnetic moments in the presence of Hund exchange in the two-band Hubbard model. We found that the formation of local magnetic moments starts below the orbital Kondo temperature $T_{\rm K, orb}$. In agreement with previous studies, this temperature scale is much higher than the Kondo temperature $T_{\rm K}$, which opens broad temperature window for the LMM formation.
In essence, it demonstrates the beginning of the formation of the LMM, conditioned by Hund correlations. At half filling we find the phase diagram, which looks similarly to the single-orbital case \cite{Our1,Our2}, except that narrow part of the QP region is found also above the insulating phase. 

In conrast to the single-orbital case, we find that preformed local moment (PLM) regime is preserved with doping. Another important feature obtained is that the temperature of screening of local magnetic moments and Kondo temperature $T_{\rm K}$ changes only slightly with doping, as does the orbital Kondo temperature $T_{K,{\rm orb}}$, discussed previously. The wide region of formation of local magnetic moments is preserved at finite doping, in contrast to the single-band model, due to contributions from the inter-orbital interactions. 

The frequency dependence of spectral functions confirms presence of the region of LMM formation, where the spectral function is suppressed at the Fermi level, while the spectral weight is redistributed to the energy of $|\sigma,\sigma\rangle$ states. We also note the presence of the LMM in the unscreened state even with significant doping, which is absent for the single-band Hubbard model. With larger doping, the local magnetic moments are  destroyed.

In the present paper, we use the Ising form of Hund exchange, which is often considered in the first principle DFT+DMFT calculations. Consideration of SU(2) symmetric Hund exchange has to be performed in future. We also do not consider magnetic correlations due to the long-range order in the ground state. In this respect, the results are applicable to frustrated lattices. Consideration of the effect of local magnetic moments on the possibility of long-range magnetic order requires considering non-local cluster or diagrammatic extensions of dynamical mean field theory. 

More generally, the results of this paper can be further used for {the} description of materials with almost formed local moments, such as Hund's metals. The consideration of the effect of local magnetic moments on the possibility of unconventional superconductivity\cite{WernerAndHoshino1, WernerAndHoshino3, WernerAndHoshino4, High-Tc1} is also of certain interest.


\vspace{-0.4cm}

\section{Acknowledgements}
The work (DMFT calculations, analysis of the results) is supported by RSF grant 24-12-00186. Development and adaptation of CT-QMC software is supported within the theme ``Quant" 122021000038-7 of Ministry of Science and Higher Education of the Russian Federation.


\appendix

\renewcommand\thefigure{B\arabic{figure}}
\renewcommand\thetable{A\arabic{table}}
\setcounter{equation}{0}
\setcounter{figure}{0}
\setcounter{table}{0}

 \begin{table*}[t]
\renewcommand{\arraystretch}{1.5}
\begin{tabular}{ |m{2.5cm}|P{2.7cm}|P{3cm}|P{2.8cm}|P{3cm}|P{2.1cm}| }
\hline
\hline
\quad Configuration & State energy $E$& $E_1=E(U'=U-2J)$ & $E_2=E_1(\mu_{HF})$ & $E_3=E_2-E_{2,\ket{\uparrow, \uparrow}}$ & $E_3(J=U/4)$ \\
\hline
\hline
\quad $\ket{\uparrow, \uparrow} , \ket{\downarrow, \downarrow}$ & $U'-J-2\mu$ & $U-3J-2\mu$ & $-2U+2J$ & $0$ & $0$\\
\hline
\quad $\ket{\uparrow, \downarrow}, \ket{\downarrow, \uparrow}$ & $U'-2\mu$ & $U-2J-2\mu$ & $-2U+3J$ & $J$ & $\frac{U}{4}$ \\
\hline
\quad $\ket{\uparrow \downarrow, 0}, \ket{0, \uparrow \downarrow}$ & $U-2\mu$ & $U-2\mu$ & $-2U+5J$ & $3J$ & $\frac{3U}{4}$ \\
\hline
\quad $\ket{\uparrow \downarrow, \uparrow}, \ket{\uparrow \downarrow, \downarrow},$ & $U+2U'-$ & \multirow{2}{*}{$3U-5J-3\mu$} & & & \\
\quad $\ket{\uparrow, \uparrow \downarrow}, \ket{\downarrow, \uparrow \downarrow}$ & $-J-3\mu$ & & \multirow{2}{*}{$-\frac{3U-5J}{2}$} & \multirow{2}{*}{$\frac{U+J}{2}$} & \multirow{2}{*}{$\frac{5U}{8}$} \\
\cline{1-3}
\quad $\ket{\uparrow, 0}, \ket{\downarrow, 0},$ & \multirow{2}{*}{$-\mu$} & \multirow{2}{*}{$-\mu$} & & & \\
\quad $\ket{0, \uparrow}, \ket{0, \downarrow}$ & & & & & \\
\hline
\quad \multirow{1}{*}{$\ket{\uparrow \downarrow, \uparrow \downarrow}$} & $2U+4U'-2J-4\mu$ & \multirow{1}{*}{$6U-10J-4\mu$} & \multirow{2}{*}{$0$} & \multirow{2}{*}{$2U - 2J$} & \multirow{2}{*}{$\frac{3U}{2}$} \\
\cline{1-3}
\quad $\ket{0, 0}$ & $0$ & $0$ & & & \\
\hline
\end{tabular}
\caption{Energies of various states in the atomic limit; $\mu_{HF}=(3U-5J)/2$ is the chemical potential, corresponding to half filling.}
\label{TableStates}
\renewcommand{\arraystretch}{1}
\end{table*}


\section{Green's function and excitation energies in the atomic limit}

 We have according to Lehmann's representation
\begin{align}
    G_{l\sigma}&(i\nu_n) = -\int _0^\beta d\tau e^{i\nu_n \tau}\langle c_{l\sigma}(\tau)c^{\dagger}_{l\sigma}(0) \rangle\notag\\
    &=- \frac{1}{Z} \sum_{m n} \int_0^\beta d\tau e^{(i\nu_n + E_m - E_n)\tau - \beta E_m} |\langle m | c_{l\sigma} | n \rangle |^2 \notag\\
    &= \frac{1}{Z} \sum_{m n} \frac{|\langle m| c_{l\sigma} | n \rangle |^2}{i\nu_n + E_m - E_n} (e^{- \beta E_m} + e^{- \beta E_n})
\end{align}
where the summation over $n$ and $m$ is performed over the states listed in Table \ref{TableStates}, $E_m$ are the respective energies, and $Z = \sum_m \exp({-\beta E_m})$ is the partition function. We therefore obtain 
\begin{eqnarray}
    G_{l\sigma}(i\nu_n) 
    &=&\frac{n_{0,\sigma}+n_{0,0}}{i\nu_n+\Delta E_1} + \frac{n_{\sigma,\sigma}+n_{\sigma,0}}{i\nu_n+\Delta E_2} \notag\\
    &+& \frac{n_{-\sigma,\sigma}+n_{-\sigma,0}}{i\nu_n+\Delta E_3} + \frac{n_{\uparrow \downarrow,0}+n_{-\sigma,0}}{i\nu_n-\Delta E_4} \notag\\
    &+& \frac{n_{\sigma, \uparrow \downarrow}+n_{\sigma,-\sigma}}{i\nu_n-\Delta E_3} + \frac{n_{\uparrow \downarrow,-\sigma}+n_{-\sigma,-\sigma}}{i\nu_n-\Delta E_2} \notag\\
    &+& \frac{n_{\uparrow \downarrow,\sigma}+n_{\uparrow \downarrow,0}}{i\nu_n+\Delta E_4} + \frac{n_{\uparrow \downarrow,\uparrow \downarrow}+n_{\uparrow \downarrow,-\sigma}}{i\nu_n-\Delta E_1}, 
\end{eqnarray}
where $n_m=\exp(-\beta E_m)/Z$ and
\begin{align} \label{eq:ex_energies}
    \Delta E_1 &= E_{\ket{0, 0}; \ket{\uparrow \downarrow, \uparrow \downarrow}}-E_{\ket{\sigma, 0}; \ket{\sigma, \uparrow \downarrow}}=\frac{3U-5J}{2} = \frac{7}{8}U, \notag\\ 
    \Delta E_2 &= E_{\ket{\sigma, 0}; \ket{\sigma, \uparrow \downarrow}}-E_{\ket{\sigma, \sigma}}=\frac{U+J}{2} = \frac{5}{8}U, \notag\\
    \Delta E_3 &= E_{\ket{\sigma, 0}; \ket{\sigma, \uparrow \downarrow}}-E_{\ket{\sigma, -\sigma}}=\frac{U-J}{2} = \frac{3}{8}U, \notag\\
    \Delta E_4 &= E_{\ket{\uparrow \downarrow, 0}}-E_{\ket{\sigma, 0}; \ket{\sigma, \uparrow \downarrow}}=\frac{5J-U}{2} = \frac{U}{8}.
\end{align}
(last equality in each line corresponds to the case $J = U/4$).



In the limit $T \rightarrow 0$ ($\beta \rightarrow \infty$) we have the only non-zero occupation numbers $n_{\sigma,\sigma}=1/2$ and therefore obtain the expression for the Green's function:
\begin{eqnarray}
    G_{l\sigma}(\nu) &=& \frac{1}{2}\left( \frac{1}{\nu + \frac{U+J}{2}} + \frac{1}{\nu - \frac{U+J}{2}} \right).
\end{eqnarray}

Further analysis of spectral functions in the presence of the hopping can be performed e.g. within the Hubbard-I approximation and broadens peaks, corresponding to Hubbard subbands. In DMFT analysis the respective peaks are further broadened 
by quasiparticle damping, originating from temperature and interaction effects.


\section{Determination of the orbital Kondo temperature from frequency dependence of the orbital susceptibility}
\label{TKorb1}

\begin{figure}[b]
\includegraphics[width=1.0\linewidth]{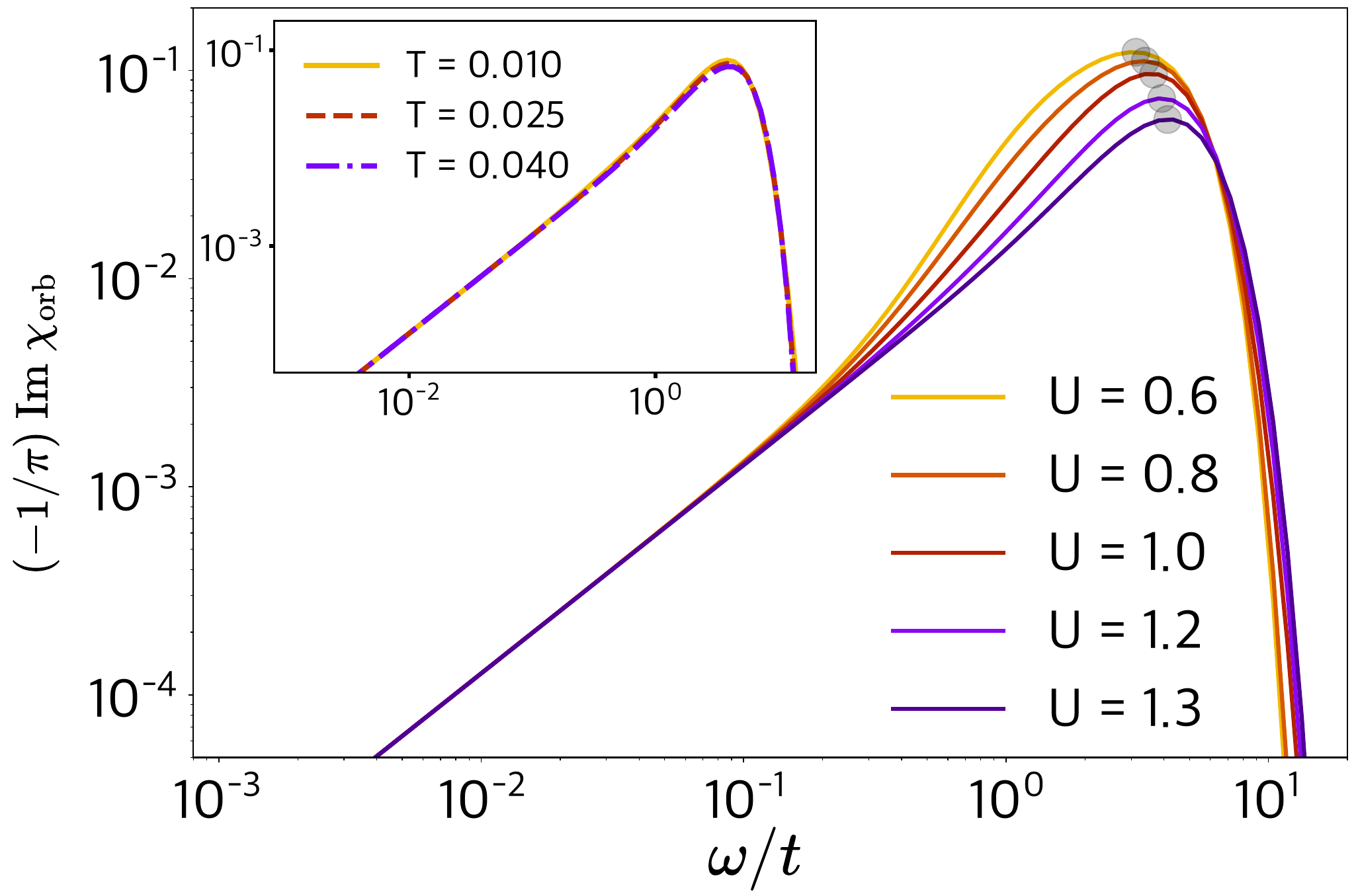}
\caption{(Color online). Main plot: Imaginary part  of orbital susceptibility at the real frequency axis at $T = 0.025$ and various values of interaction $U$. Maxima of $\rm Im \ \chi_{orb}$ are indicated by shaded circles ($T^\omega_{\rm K,orb}$). Inset: imaginary part of orbital susceptibility at $U = 1.2$ and various $T$.}
\label{fig:chi_o_stadler}
\end{figure}

{Following Ref. \cite{Hund3}, we consider (apart from the temperature dependence of orbital susceptibilities) the possibility to extract orbital Kondo temperatures from the frequency dependence of imaginary part of the orbital susceptibilities. We use {analytical continuation of the dynamic orbital susceptibility (\ref{eq:chi_orb}) to real frequencies by maximum entropy method \cite{ana_cont}. Typical results of such calculation are shown in Fig. \ref{fig:chi_o_stadler}. We consider orbital susceptibility at temperatures low enough that peak positions are not affected, as can be seen in the inset of Fig. \ref{fig:chi_o_stadler}. According to Ref. \cite{Hund3}, the peak of the susceptibility corresponds to a characteristic energy scale, which is denoted by $T^\omega_{\rm K,orb}$ in the main text. The linear behavior of the spectral density at low frequencies is the characteristic for the Fermi liquid. The positions of the maxima are proportional to the Kondo temperatures $T_{\rm K,orb}$, determined from the temperature dependencies of the orbital susceptibility, as described in the discussion of Fig. \ref{fig:phase_diagram_hund}} of the main text.}


\appendix
\renewcommand\theequation{A\arabic{equation}}
\renewcommand\thefigure{A\arabic{figure}}
\renewcommand\thetable{A\arabic{table}}
\setcounter{equation}{0}
\setcounter{figure}{0}
\setcounter{table}{0}




\begin{thebibliography}{1}

\bibitem{Freezing} P. Werner, E. Gull, M. Troyer, and A. J. Millis, Spin Freezing Transition and Non-Fermi-Liquid Self-Energy in a Three-Orbital Model, \href{https://doi.org/10.1103/PhysRevLett.101.166405}{Phys. Rev. Lett. {\bf 101}, 166405 (2008)}.
\bibitem{Pnictides} Z. P. Yin, K. Haule, and G. Kotliar, Kinetic frustration and the nature of the magnetic and paramagnetic states in iron pnictides and iron chalcogenides, \href{https://doi.org/10.1038/nmat3120}{Nature Materials {\bf 10}, 932 (2011)}.
\bibitem{Medici} L. de' Medici, Hund’s coupling and its key role in tuning multiorbital correlations, \href{https://doi.org/10.1103/PhysRevB.83.205112}{Phys. Rev. B {\bf 83}, 205112 (2011)}.
\bibitem{Hund1} L. de' Medici, J. Mravlje, A. Georges, Janus-Faced Influence of Hund’s Rule Coupling in Strongly Correlated Materials, \href{https://doi.org/10.1103/PhysRevLett.107.256401}{Phys. Rev. Lett. {\bf 107}, 256401 (2011)}.
\bibitem{Pnictides1}
A. Georges, L. de' Medici, and J. Mravlje, Strong Correlations from Hund’s Coupling, \href{https://doi.org/10.1146/annurev-conmatphys-020911-125045}{Ann. Rev. Cond. Matt. Phys. {\bf 4}, 137 (2013)}.
\bibitem{Delft}
K. M. Stadler, Z. P. Yin, J. von Delft, G. Kotliar, and A. Weichselbaum, Dynamical Mean-Field Theory Plus Numerical Renormalization-Group Study of Spin-Orbital Separation in a Three-Band Hund Metal, \href{https://doi.org/10.1103/PhysRevLett.115.136401}{Phys. Rev. Lett. {\bf 115}, 136401 (2015)}.
\bibitem{Hund2} L. de' Medici, Hund's metals, explained, in ``The Physics of Correlated Insulators, Metals, and Superconductors Modeling and Simulation", Schriften des Forschungszentrums Jülich
Reihe Modeling and Simulation, Vol. 7, Forschungszentrum Juelich, 2017, ISBN 978-3-95806-224-5, \url{https://www.cond-mat.de/events/correl17/manuscripts/correl17.pdf}. 
\bibitem{Hund3} K. M. Stadler, G. Kotliar, A. Weichselbaum, J. von Delft, Hundness versus Mottness in a three-band Hubbard–Hund model: On the origin of strong correlations in Hund metals, \href{https://doi.org/10.1016/j.aop.2018.10.017}{Annals of Physics {\bf 405}, 365 (2019)}.
\bibitem{Hund4} E. Walter, K. M. Stadler, S. -S. B. Lee, Y. Wang, G. Kotliar, A. Weichselbaum, J. von Delft, Uncovering Non-Fermi-Liquid Behavior in Hund Metals: Conformal Field Theory Analysis of an $\mathrm{SU}(2)\ifmmode\times\else\texttimes\fi{}\mathrm{SU}(3)$ Spin-Orbital Kondo Model, \href{https://doi.org/10.1103/PhysRevX.10.031052}{Phys. Rev. X {\bf 10}, 031052 (2020)}.
\bibitem{Hund5} K. M. Stadler, G. Kotliar, S. -S. B. Lee, A. Weichselbaum, J. von Delft, Differentiating Hund from Mott physics in a three-band Hubbard-Hund model: Temperature dependence of spectral, transport, and thermodynamic properties, \href{https://doi.org/10.1103/PhysRevB.104.115107}{Phys. Rev. B {\bf 104}, 115107 (2021)}.
\bibitem{Pnic1} L. de' Medici, Weak and strong electronic correlations in Fe superconductors, in ``Iron-based Superconductivity", Springer Series in Materials Science, Vol. 211, pp. 409-441 (2015), \url{https://doi.org/10.1007/978-3-319-11254-1_11}
\bibitem{Pnic2} P. V. Arribi, L. de' Medici, Hund's metal crossover and superconductivity in the 111 family of iron-based superconductors, \href{https://doi.org/10.1103/PhysRevB.104.125130}{Phys. Rev. B {\bf 104}, 125130 (2021)}.
\bibitem{Hund1x} J. Mravlje, M. Aichhorn, T. Miyake, K. Haule, G. Kotliar, and A. Georges, Coherence-Incoherence Crossover and the Mass-Renormalization Puzzles in ${\mathrm{Sr}}_{2}{\mathrm{RuO}}_{4}$, \href{https://doi.org/10.1103/PhysRevLett.106.096401}{Phys. Rev. Lett. {\bf 106}, 096401 (2011)}.
\bibitem{Hund2x} J. Mravlje and A. Georges, Thermopower and Entropy: Lessons from ${\mathrm{Sr}}_{2}{\mathrm{RuO}}_{4}$, \href{https://doi.org/10.1103/PhysRevLett.117.036401}{Phys. Rev. Lett. {\bf 117}, 036401 (2016)}.
\bibitem{OurFe1}  A. A. Katanin, A. I. Poteryaev, A. V. Efremov, A. O. Shorikov, S. L. Skornyakov, M. A. Korotin, and V. I. Anisimov, Orbital-selective formation of local moments in $\ensuremath{\alpha}$-iron: First-principles route to an effective model, \href{https://doi.org/10.1103/PhysRevB.81.045117}{Phys. Rev. B {\bf 81}, 045117 (2010)}.
\bibitem{OurFeGamma} P. A. Igoshev, 
A. V. Efremov, A. I. Poteryaev, A. A. Katanin, and V. I. Anisimov, Magnetic fluctuations and effective magnetic moments in $\ensuremath{\gamma}$-iron due to electronic structure peculiarities, \href{https://doi.org/10.1103/PhysRevB.88.155120}{Phys. Rev. B {\bf 88}, 155120 (2013)}.
\bibitem{Hausoel} A. Hausoel, M. Karolak, E. Sasioglu, A. Lichtenstein, K. Held, A. Katanin, A. Toschi, and G. Sangiovanni, Local magnetic moments in iron and nickel at ambient and Earth’s core conditions, \href{https://doi.org/10.1038/ncomms16062}{Nat. Commun. {\bf 8}, 16062 (2017)}.
\bibitem{OurIron} A. S. Belozerov, A. A. Katanin, and V. I. Anisimov, Momentum-dependent susceptibilities and magnetic exchange in bcc iron from supercell dynamical mean-field theory calculations, \href{https://doi.org/10.1103/PhysRevB.96.075108}{Phys. Rev. B {\bf 96}, 075108 (2017)}.
\bibitem{OurV}  A. S. Belozerov, A. A. Katanin, V. I. Anisimov, Transition from Pauli paramagnetism to Curie-Weiss behavior in vanadium, \href{https://doi.org/10.1103/PhysRevB.107.035116}{Phys. Rev. B {\bf 107}, 035116, (2023)}.
\bibitem{OurFeNi} A. S. Belozerov, A. A. Katanin, and V. I. Anisimov, Electronic correlation effects and local magnetic moments in ${\rm L1_0}$ phase of FeNi, \href{https://doi.org/10.1088/1361-648X/ab9566}{J. Phys.: Condens. Matter {\bf 32}, 385601 (2020)}.
\bibitem{High-Tc} M. Ferrero, P. S. Cornaglia, L. De Leo, O. Parcollet, G. Kotliar, and A. Georges, Pseudogap opening and formation of Fermi arcs as an orbital-selective Mott transition in momentum space, \href{https://doi.org/10.1103/PhysRevB.80.064501}{Phys. Rev. B {\bf 80}, 064501 (2009)}.
\bibitem{High-Tc1} P. Werner, S. Hoshino, and H. Shinaoka, Spin-freezing perspective on cuprates, \href{https://doi.org/10.1103/PhysRevB.94.245134}{Phys. Rev. B {\bf 94}, 245134 (2016)}.
\bibitem{Fanfarillo} L. Fanfarillo and E. Bascones, Electronic correlations in Hund metals, \href{https://doi.org/10.1103/PhysRevB.92.075136}{Phys. Rev. B {\bf 92}, 075136 (2015)}.
\bibitem{Deng} X. Deng, K. M. Stadler, K. Haule, A. Weichselbaum, J. von Delft, and G. Kotliar, Signatures of Mottness and Hundness in archetypal correlated metals, \href{https://doi.org/10.1038/s41467-019-10257-2}{Nature Communications {\bf 10}, 2721 (2019)}; X. Deng, K. M. Stadler, K. Haule, S. -S. B. Lee, A. Weichselbaum, J. von Delft, and G. Kotliar, Reply to: Extracting Kondo temperature of strongly-correlated systems from the inverse local magnetic susceptibility, \href{https://doi.org/10.1038/s41467-021-21643-0}{Nat. Commun. {\bf 12}, 1445 (2021)}.
\bibitem{Toschi} P. Chalupa, T. Sch\"afer, M. Reitner, D. Springer, S. Andergassen, and A. Toschi, Fingerprints of the Local Moment Formation and its Kondo Screening in the Generalized Susceptibilities of Many-Electron Problems, \href{https://doi.org/10.1103/PhysRevLett.126.056403}{Phys. Rev. Lett. {\bf 126}, 056403 (2021)}.
\bibitem{Katsnelson} E. A. Stepanov, S. Brener, V. Harkov, M. I. Katsnelson, and A. I. Lichtenstein, Spin dynamics of itinerant electrons: Local magnetic moment formation and Berry phase, \href{https://doi.org/10.1103/PhysRevB.105.155151}{Phys. Rev. B {\bf 105}, 155151 (2022)}.
\bibitem{Our1} T. B. Mazitov, A. A. Katanin, Local magnetic moment formation and Kondo screening in the half-filled single-band Hubbard model, \href{https://doi.org/10.1103/PhysRevB.105.L081111}{Phys. Rev. B {\bf 105}, L081111 (2022)}.
\bibitem{Our2} T. B. Mazitov, A. A. Katanin, Effect of local magnetic moments on spectral properties and resistivity near interaction- and doping-induced Mott transitions, \href{https://doi.org/10.1103/PhysRevB.106.205148}{Phys. Rev. B {\bf 106}, 205148 (2022)}.
\bibitem{Toschi_new} S. Adler, F. Krien, P. Chalupa-Gantner, G. Sangiovanni, and A. Toschi, Non-perturbative intertwining between spin and charge correlations: A ``smoking gun'' single-boson-exchange result, \href{https://doi.org/10.21468/SciPostPhys.16.2.054}{SciPost Phys. {\bf 16}, 054 (2024)}.
\bibitem{Pruschke} Th. Pruschke and R. Bulla, Hund’s coupling and the metal-insulator transition in the two-band Hubbard model, \href{https://doi.org/10.1140/epjb/e2005-00117-4}{Eur. Phys. J. B {\bf 44}, 217 (2005)}.
\bibitem{2band1} J. Steinbauer, L. de' Medici, and S. Biermann, Doping-driven metal-insulator transition in correlated electron systems with strong Hund's exchange coupling, \href{https://doi.org/10.1103/PhysRevB.100.085104}{Phys. Rev. B {\bf 100}, 085104 (2019)}.
\bibitem{Anisimov} D. Y. Novoselov, D. M. Korotin, A. O. Shorikov, and V. I. Anisimov, Charge and spin degrees of freedom in strongly correlated systems: Mott states opposite Hund's metals, \href{https://doi.org/10.1088/1361-648X/ab7600}{J. Phys.: Condens. Matter {\bf 32}, 235601 (2020)}.
\bibitem{Werner} K. Steiner, S. Hoshino, Y. Nomura, and P. Werner, Long-range orders and spin/orbital freezing in the two-band Hubbard model, \href{https://doi.org/10.1103/PhysRevB.94.075107}{Phys. Rev. B {\bf 94}, 075107 (2016)}.
\bibitem{2orb1} S. Ryee, S. Choi, M. J. Han, Hund Physics Landscape of Two-Orbital Systems, \href{https://doi.org/10.1103/PhysRevLett.126.206401}{Phys. Rev. Lett. {\bf 126}, 206401 (2021)}; S. Ryee, S. Choi, M. J. Han, Frozen spin ratio and the detection of Hund correlations, \href{https://doi.org/10.1103/PhysRevResearch.5.033134}{Phys. Rev. Research {\bf 5}, 033134 (2023)}.
\bibitem{DMFT} A. Georges, G. Kotliar, W. Krauth, and M. Rozenberg, Dynamical mean-field theory of strongly correlated fermion systems and the limit of infinite dimensions, \href{https://doi.org/10.1103/RevModPhys.68.13}{Rev. Mod. Phys. {\bf 68}, 13 (1996)}.
\bibitem{LocDMFT1} M. Jarrell and Th. Pruschke, Magnetic and dynamic properties of the Hubbard model in infinite dimensions, \href{https://doi.org/10.1007/BF02198153}{Z. Phys. B {\bf 90}, 187 (1993)}; Anomalous properties of the Hubbard model in infinite dimensions, \href{https://doi.org/10.1103/PhysRevB.49.1458}{Phys. Rev. B {\bf 49}, 1458 (1994)}.
\bibitem{ToschiAnd1} P. Chalupa, P. Gunacker, T. Sch\"afer, K. Held, and A. Toschi, Divergences of the irreducible vertex functions in correlated metallic systems: Insights from the Anderson impurity model, \href{https://doi.org/10.1103/PhysRevB.97.245136}{Phys. Rev. B {\bf 97}, 245136 (2018)}.
\bibitem{Comment} A. A. Katanin, Extracting Kondo temperature of strongly-correlated systems from the inverse local magnetic susceptibility, \href{https://doi.org/10.1038/s41467-021-21641-2}{Nat. Commun. {\bf 12}, 1433 (2021)}.
\bibitem{cthyb1} P. Werner, A. Comanac, L. de' Medici, M. Troyer, and A. J. Millis, Continuous-Time Solver for Quantum Impurity Models, \href{https://doi.org/10.1103/PhysRevLett.97.076405}{Phys. Rev. Lett {\bf 97}, 076405 (2006)}.
\bibitem{cthyb2} P. Werner and A. J. Millis, Hybridization expansion impurity solver: General formulation and application to Kondo lattice and two-orbital models, \href{https://doi.org/10.1103/PhysRevB.74.155107}{Phys. Rev. B {\bf 74}, 155107 (2006)}.
\bibitem{cthyb3} E. Gull, A. J. Millis, A. I. Lichtenstein, A. N. Rubtsov, M. Troyer, and P. Werner, Continuous-time Monte Carlo methods for quantum impurity models, \href{https://doi.org/10.1103/RevModPhys.83.349}{Rev. Mod. Phys. {\bf 83}, 349 (2011)}.
\bibitem{cthyb4} H. Hafermann, K. R. Patton, and P. Werner, Improved estimators for the self-energy and vertex function in hybridization-expansion continuous-time quantum Monte Carlo simulations, \href{https://doi.org/10.1103/PhysRevB.85.205106}{Phys. Rev. B {\bf 85}, 205106 (2012)}.
\bibitem{cthyb5} H. Hafermann, Self-energy and vertex functions from hybridization-expansion continuous-time quantum Monte Carlo for impurity models with retarded interaction, \href{https://doi.org/10.1103/PhysRevB.89.235128}{Phys. Rev. B {\bf 89}, 235128 (2014)}.
\bibitem{iQIST} L. Huang, Y. Wang, Z. Y. Meng, L. Du, P. Werner, and X. Dai, iQIST: An open source continuous-time quantum Monte Carlo impurity solver toolkit, \href{https://doi.org/10.1016/j.cpc.2015.04.020}{Comput. Phys. Commun. \textbf{195}, 140 (2015)}; L. Huang, 
iQIST v0.7: An open source continuous-time quantum Monte Carlo impurity solver toolkit, \href{https://doi.org/10.1016/j.cpc.2017.08.026}{\textit{ibid.} \textbf{221}, 423 (2017)}.
\bibitem{iQISTNote} {The integrals over $\tau$ in charge correlators and spin correlators are estimated as sums over CT-QMC imaginary time segments.}  
\bibitem{ana_cont} J. Kaufmann and K. Held, ana\_cont: Python package for analytic continuation, \href{https://doi.org/10.1016/j.cpc.2022.108519}{Comput. Phys. Commun. {\bf 282}, 108519 (2023)}; \url{https://github.com/josefkaufmann/ana\_cont}.


\bibitem{WernerAndHoshino3} S. Hoshino and P. Werner, Superconductivity from Emerging Magnetic Moments, \href{https://doi.org/10.1103/PhysRevLett.115.247001}{Phys. Rev. Lett. {\bf 115}, 247001 (2015)}.
\bibitem{MediciTwoBand} M. Chatzieleftheriou, A. Kowalski, M. Berović, A. Amaricci, M. Capone, L. De Leo, G. Sangiovanni, and L. de' Medici, Mott Quantum Critical Points at Finite Doping, \href{https://doi.org/10.1103/PhysRevLett.130.066401}{Phys. Rev. Lett {\bf 130}, 066401 (2023)}.
\bibitem{WernerAndHoshino4} P. Werner, A. J. Kim, and S. Hoshino, Spin-freezing and the Sachdev-Ye model, \href{https://doi.org/10.1209/0295-5075/124/57002}{Europhys. Lett. {\bf 124}, 57002 (2018)}.
\bibitem{WernerAndHoshino1} P. Werner and S. Hoshino, Nickelate superconductors: Multiorbital nature and spin freezing, \href{https://doi.org/10.1103/PhysRevB.101.041104}{Phys. Rev. B {\bf 101}, 041104(R) (2020)}.

















%
\end{thebibliography}
\end{document}